\documentclass[preprint,amsmath,amssymb,showkeys,superscriptaddress,showpacs,citeautoscript,floatfix]{revtex4-1}
\usepackage{graphicx}
\usepackage{dcolumn}
\usepackage{bm}
\usepackage{amsmath}
\usepackage{amssymb}
\usepackage[]{units}
\usepackage{subcaption}
\usepackage{booktabs}
\usepackage{lineno}
\begin{document}

\title{Suppression of material transfer at contacting surfaces: \\The effect of adsorbates on Al/TiN and Cu/diamond interfaces \\from first-principles calculations}

\author{Gregor Feldbauer}
\email{gregor.feldbauer@tuhh.de}
%\phone{+49 (0)40 428783644}
%\fax{+49 (0)40 428782647}
\affiliation{Institute of Advanced Ceramics, Hamburg University of Technology, Denickestra\ss e 15, 21073 Hamburg, Germany}
\affiliation{Institute of Applied Physics, Vienna University of Technology, Wiedner Hauptstra\ss e 8, 1040 Vienna, Austria}
\affiliation{Austrian Center of Competence for Tribology, AC2T research GmbH, Viktor-Kaplan-Stra\ss e 2, 2700 Wiener Neustadt, Austria}
\author{Michael Wolloch}
\affiliation{Institute of Applied Physics, Vienna University of Technology, Wiedner Hauptstra\ss e 8, 1040 Vienna, Austria}
\affiliation{Austrian Center of Competence for Tribology, AC2T research GmbH, Viktor-Kaplan-Stra\ss e 2, 2700 Wiener Neustadt, Austria}
\affiliation{Department of Physics, Informatics and Mathematics, University of Modena and Reggio Emilia, Via Campi 213/A, 41125 Modena, Italy} 
\author{Pedro O. Bedolla}
\affiliation{Institute of Applied Physics, Vienna University of Technology, Wiedner Hauptstra\ss e 8, 1040 Vienna, Austria}
\affiliation{Austrian Center of Competence for Tribology, AC2T research GmbH, Viktor-Kaplan-Stra\ss e 2, 2700 Wiener Neustadt, Austria}
\author{Josef Redinger}
\affiliation{Institute of Applied Physics, Vienna University of Technology, Wiedner Hauptstra\ss e 8, 1040 Vienna, Austria}
\author{Andr\'as Vernes}
\affiliation{Institute of Applied Physics, Vienna University of Technology, Wiedner Hauptstra\ss e 8, 1040 Vienna, Austria}
\affiliation{Austrian Center of Competence for Tribology, AC2T research GmbH, Viktor-Kaplan-Stra\ss e 2, 2700 Wiener Neustadt, Austria}
\author{Peter Mohn}
\affiliation{Institute of Applied Physics, Vienna University of Technology, Wiedner Hauptstra\ss e 8, 1040 Vienna, Austria}

\begin{abstract}
The effect of monolayers of oxygen (O) and hydrogen (H) on the possibility of material transfer at aluminium/titanium nitride (Al/TiN) and copper/diamond (Cu/C$_{\text{dia}}$) interfaces, respectively, were investigated within the framework of density functional theory (DFT). To this end the approach, contact, and subsequent separation of two atomically flat surfaces consisting of the aforementioned pairs of materials were simulated. These calculations were performed for the clean as well as oxygenated and hydrogenated Al and C$_{\text{dia}}$ surfaces, respectively. Various contact configurations were considered by studying several lateral arrangements of the involved surfaces at the interface. Material transfer is typically possible at interfaces between the investigated clean surfaces; however, the addition of O to the Al and H to the C$_{\text{dia}}$ surfaces was found to hinder material transfer. This passivation occurs because of a significant reduction of the adhesion energy at the examined interfaces, which can be explained by the distinct bonding situations.

\end{abstract}

%71.15.Mb, Density functional theory, local density approximation, gradient and other corrections
%62.20.Qp, Tribology of solids, Friction, tribology, and hardness
%81.07.Lk, Materials: nanoscale materials: nanocontacts
%31.15.E-, Electronic structure: ab initio calculations: density-functional theory


\maketitle

\newpage

\section{Introduction}\label{sec:intro}

Understanding atomistic phenomena at contacting surfaces is fundamental to the improvement of many modern applications, ranging from experimental methods like atomic or friction force microscopy (AFM/FFM)~\cite{binnig_atomic_1986,  bennewitz_friction_2005} or nanoindentation~\cite{fischer-cripps_nanoindentation_2004} to nanotechnologies employed, for example, in nano-/microelectromechanical-systems (NEMS/MEMS)~\cite{kim_nanotribology_2007, bhushan_nanotribology_2008}. Particularly, the overall performance of NEMS depends sensitively on the interfacial processes between contacting materials. Moreover, nanotribological processes, like nanoscale wear~\cite{bhushan_nanotribolgy_1995, gnecco_abrasive_2002, gotsmann_atomistic_2008, bhaskaran_ultralow_2010, jacobs_nanoscale_2013}, are not yet understood comprehensively because of their highly complex nature~\cite{kim_nano-scale_2012}. Thus, detailed studies of relevant interfaces are necessary for a better description of such processes. 

Heterointerfaces between metals and counterparts like ceramics~\cite{howe_bonding_1993} or diamond~\cite{guo_adhesion_2010} are of high technological as well as fundamental interest because they combine benefits of both types of the involved material classes, such as high thermal stability, degradation resistance, and interesting mechanical properties~\cite{johansson_electronic_1995, wang_copper/diamond_2001}. Such interfaces are used for various applications ranging from protective coatings to communication devices and (nano)electronics~\cite{ruhle_preface_1992, kawarada_hydrogen-terminated_1996}. Two distinct, technologically highly relevant material pairings are investigated in this paper, namely the metal-ceramic Al/TiN interface and the metal-insulator Cu/diamond (C$_{\text{dia}}$) interface. These interfaces are conceived as contacts between hard and soft materials.

In reality, however, surfaces are usually not pristine. For example, when aluminum is exposed to air a thin oxide film is formed at the Al surface. This passivation prevents further oxidation and results in an excellent corrosion resistance of the material~\cite{vargel_corrosion_2004}. The typical thickness of such a film is up to \unit[50]{nm}. As a first approximation on the route towards such exceedingly complex interfaces, the effect of a monolayer of oxygen covering Al surfaces will be discussed in this work. The adsorption of oxygen atoms at an Al surface also constitutes the initial step during the formation of surface oxides. For C$_{\text{dia}}$ it is important to consider the effect of hydrogen atoms, because they play a crucial role in the chemical vapour deposition (CVD) growth of diamond and they passivate dangling bonds at diamond surfaces.~\cite{kawarada_hydrogen-terminated_1996, wang_copper/diamond_2001} Thus, the influence of a monolayer of hydrogen at diamond surfaces is investigated here.

Beginning in the 1980s, classical molecular dynamics (MD) simulations have become a standard tool to computationally investigate nanotribological phenomena, see, e.g., Refs.~\cite{thompson_simulations_1989, bhushan_computer_2000, kenny_molecular_2005, david_schall_molecular_2007, szlufarska_recent_2008, vernes_three-term_2012, eder_applicability_2015, eder_evolution_2015, eder_methods_2016}. During the last decade, additionally, the use of density functional theory (DFT) calculations has been introduced in nanotribology, see, e.g., Refs.~\cite{zhong_first-principles_1990, dag_atomic_2004,  garvey_shear_2011, zilibotti_ab_2011, cahangirov_frictional_2012, kwon_enhanced_2012, bedolla_density_2014, wolloch_ab-initio_2014, feldbauer_adhesion_2015, wolloch_ab_2015}. The main advantage of DFT is the independence of empirical potentials, i.e., DFT allows for parameter-free calculations via an accurate quantum-mechanical  description of systems. On the other hand, DFT calculations are currently limited to relatively small systems of typically a few hundred atoms because of computational challenges. Thus, DFT should be seen as an extension to and not as a replacement for the more common computational tools in tribology. Since DFT calculations give very reliable results for the studied class of systems~\cite{finnis_theory_1996, lundqvist_density-functional_2001, sinnott_ceramic/metal_2003}, this method is employed here to analyse the electronic and atomic structure of the investigated interfaces, e.g., to determine adhesion energies. DFT results, such as potential-energy curves, can be utilized as a starting point for a multi-scale approach in which data is handed over to, e.g., large-scale classical MD simulations~\cite{ercolessi_interatomic_1994, jaramillo-botero_general_2014}. In the last years also quantum-classical embedding techniques have been developed and improved allowing to treat crucial parts of a system with high accuracy methods such as DFT, while most of the system is evaluated using less expensive methods.~\cite{csanyi_learn_2004, moras_atomically_2010, kermode_macroscopic_2013} Such joint approaches combined with advances in software tools and the continuously increasing available computer power promise the feasibility to study even larger and more realistic systems in the near future. 

Investigations on Al/TiN interfaces have been conducted by various research groups using experimental~\cite{avinun_nucleation_1998, chun_interfacial_2001-1, howe_bonding_1993, howe_bonding_1993-1, ernst_metal-oxide_1995} as well as theoretical~\cite{liu_first-principles_2004, song_adhesion_2006, zhang_first-principles_2007, zhang_effects_2007, yadav_first-principles_2012, yadav_first-principles_2014, feldbauer_adhesion_2015, yadav_structural_2015, li_growth_2015-1, lin_atomic_2017} methods. The role of interfacial species at Al/TiN interfaces is, however, less studied. Liu et al.~\cite{liu_first-principles_2005} and Zhang et al.~\cite{zhang_effects_2007} investigated the effects of hydrogen and Zn as well as Mg, respectively, on Al/TiN interfaces. Both computational studies concluded that the interfacial adhesion is reduced by the additional species at the interface. Here, the emphasis lies on the role of oxygen, since aluminium is usually covered by an oxide layer under ambient conditions.~\cite{vargel_corrosion_2004} Further information on oxide layers on aluminium surfaces can be found, e.g., in Refs.~\cite{jacobsen_electronic_1995, jennison_ultrathin_2000}. As a first step towards a more detailed description of Al/TiN interfaces, the Al slab is terminated by one layer of oxygen in the current work, which focuses on a possible material transfer and its ab-initio simulation. Material transfer is here defined as the detachment of atoms or atomic layers from one body and the reattachment to the counterbody during the course of one loading cycle. In this context one loading cycle describes the approach, contact and subsequent separation of two initially separated bodies.

Insights on copper/diamond interfaces with and without interfacial species have been presented by various researchers~\cite{guo_adhesion_2010, schaich_structural_1997, scholze_structure_1996, kern_atomic_1996, kern_atomic_1996-1, kern_atomic_1997, wang_copper/diamond_2001, pepper_effect_1982, zhu_study_2016}. Cu/C$_{\text{dia}}$ will be used as an abbreviation in the following. For Cu/C$_{\text{dia}}$ interfaces, Guo et al.~\cite{guo_adhesion_2010} studied the fracture behaviour by comparing the work of separation and the work of decohesion. The structure and work of separation for such interfaces including H terminated diamond was examined computationally by Wang and Smith~\cite{wang_copper/diamond_2001}. They found a strong decrease in the adhesion energy and thus the work of separation of Cu/C$_{\text{dia}}$ interfaces upon the introduction of interfacial hydrogen, but they did not investigate material transfer explicitly. Their results are in agreement with experiments performed by Pepper~\cite{pepper_effect_1982} on the frictional properties of these interfaces. Furthermore, in a recent study involving MD simulations of nanoscratching Zhu et al.~\cite{zhu_study_2016} observed that the minimum depth of material transfer at Cu surfaces can be as thin as only one atomic layer of Cu depending on the machining conditions. In this work, as for the Al/TiN interfaces, the emphasis is put on material transfer and its ab-initio simulation.

\section{Computational Details}\label{sec:comp_methods}

\subsection{Density Functional Theory Calculations}

To investigate the properties of Al/TiN interfaces and Cu/C$_{\text{dia}}$ interfaces terminated with oxygen and hydrogen, respectively, we performed  calculations of the approach and subsequent separation of two slabs forming one of the aforementioned interfaces. The focus of this paper is put on Al/TiN interfaces, while results on contacting Cu and diamond slabs are given for comparison, as a validation of the method, and a more general view of the subject. All results were calculated within the framework of DFT using the Vienna Ab initio Simulation Package (\textsc{VASP})~\cite{kresse_ab_1993, kresse_ab_1994, kresse_efficient_1996, kresse_efficiency_1996}. \textsc{VASP} uses  periodic boundary conditions and a plane-wave basis set within the projector augmented-wave (PAW) formalism~\cite{blochl_projector_1994,kresse_ultrasoft_1999}. The interaction between the ionic core and the valence electrons is mediated by the corresponding PAW pseudopotentials. The generalized gradient approximation (GGA) in the Perdew, Burke, and Ernzerhof (PBE) parametrization~\cite{perdew_generalized_1996} was used to describe the exchange and correlation (XC) functional, unless explicitly mentioned otherwise. While GGAs often underestimate binding and adhesion energies~\cite{stampfl_density-functional_1999}, the local-density approximation (LDA)~\cite{perdew_self-interaction_1981} usually overestimates these quantities~\cite{van_de_walle_correcting_1999}. Further examples but also deviations from this trend for the behaviour of different XC functionals are presented in Ref.~\cite{grabowski_ab_2007}. Therein, it is observed that LDA and GGA results on material properties often give a lower and upper bound for corresponding experimental data. Some trends exist for specific properties, for example, LDA often performs better for surface energies, while PBE gives usually better results for elastic properties and cohesive energies.~\cite{csonka_assessing_2009, wandelt_introduction_2014} No functional is, however, considered to be generally superior to the other ones. Here, calculations employing the LDA method are used for comparison and to check the consistency of the results.

Accurate total energies are guaranteed by a careful choice of the computational parameters. The \textit{k}-space integrations were performed using a \(\Gamma\)-centred \(15\times15\times1\) Monkhorst-Pack mesh~\cite{monkhorst_special_1976}. The energy cutoff for the plane-wave basis was set to \unit[800]{eV}. Both settings result in total energies converged to \unit[1]{meV/atom}. While atomic relaxations were performed employing the method of Methfessel and Paxton~\cite{methfessel_high-precision_1989} to first-order with a smearing of \unit[0.11]{eV}, the tetrahedron method with Bl\"ochl corrections~\cite{blochl_improved_1994} was used for all static calculations, i.e., calculations with fixed atomic positions. Since the method of Methfessel and Paxton should only be used for metallic systems, Gaussian smearing of \unit[0.05]{eV} was chosen for relaxations of the Cu/C$_{\text{dia}}$ interfaces. Instead of the widely used conjugate-gradient or quasi-Newton algorithms a damped molecular dynamics (MD) scheme was selected to relax the atomic structures. This choice was made because the first two algorithms showed convergence problems and they tended to remain stuck in local minima.  To obtain more accurate total energies, each converged relaxation was complemented by a follow-up static calculation. For relaxations and electronic self-consistency cycles the convergence criteria of \unit[\(10^{-5}\)]{eV} and \unit[\(10^{-6}\)]{eV}, respectively, were employed. All simulations were performed at \unit[0]{K}.

\subsection{Simulation Model}\label{subsec:sim-model}

The simulation model is described here for the case of an Al/TiN interface, using the methodology presented by the authors in an earlier publication~\cite{feldbauer_adhesion_2015}. The same route is followed in the case of Cu/C$_{\text{dia}}$ interfaces. 

\begin{figure}[hbt]
  \centering
  \begin{subfigure}[b]{0.17\linewidth}
    \includegraphics[width=\linewidth]{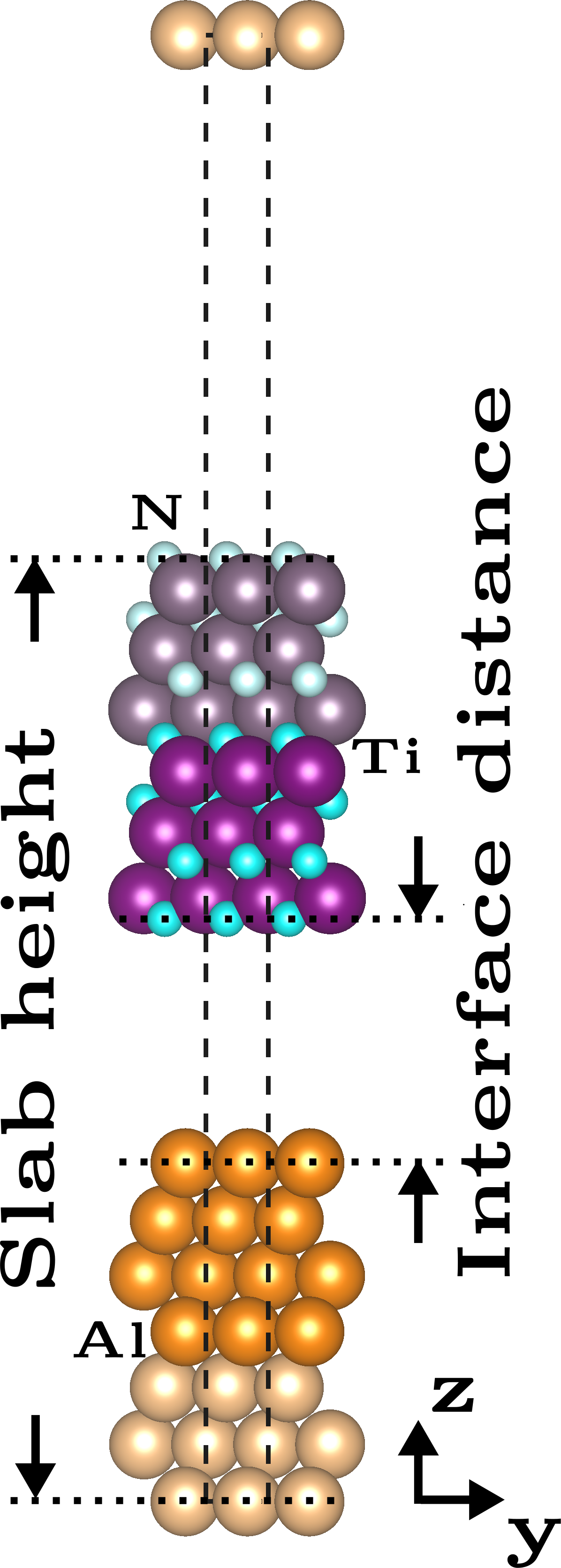}
      \caption{\\Al/TiN (hcp)}
    \label{fig:al-tin-side}
  \end{subfigure}
  \qquad
  \qquad
  \begin{subfigure}[b]{0.17\linewidth}
    \includegraphics[width=\linewidth]{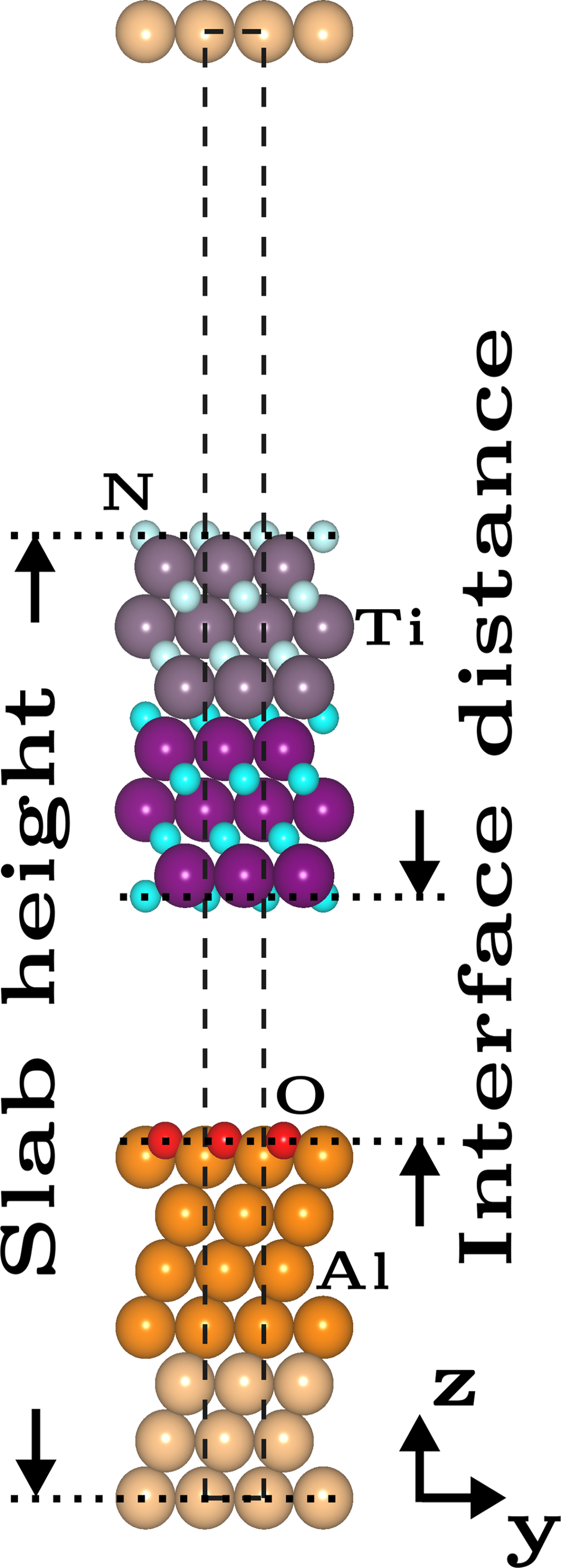}
  \caption{\\Al-O/TiN~(top)}
    \label{fig:al-o-tin-side}
  \end{subfigure}
  \caption{Side view of two (111) Al/TiN interfaces (TiN: N terminated). The Al slab is pristine in (a) and oxygenated in (b). The simulation interface cell is indicated by the dashed black lines. Orange, red, cyan, and purple identify Al, O, N, and Ti atoms, respectively. During relaxations the pale colors indicate atoms that were kept frozen, while the strong colors mark atoms that were allowed to relax.}
  \label{fig:Al-TiN-initial}
\end{figure}

To model our Al/TiN systems we constructed simulation cells from a fcc (111) Al slab at the bottom and a rock salt (111) TiN slab above (see Fig.~\ref{fig:al-tin-side}). The gap between the two slabs is called the ``interface distance'' which is measured from the top layer of the lower slab to the bottom layer of the upper slab, regardless of termination. In the following the term ``slab height'' refers to the vertical distance between the bottom Al and top TiN layers, which is the sum of the interface distance and the heights of the two slabs. To investigate the effect of oxygenation, an Al slab was terminated by oxygen at the interface (see Fig.~\ref{fig:al-o-tin-side}). In this work 1\(\times\)1 surface cells were used, which represent infinitely extended surfaces because of the periodic boundary conditions. The TiN slab consisted of a minimum of six Ti and six N layers, while the Al slab are built of seven layers. These geometric parameters are consistent with previous investigations.~\cite{feldbauer_adhesion_2015, marlo_density-functional_2000, liu_first-principles_2004, zhang_first-principles_2007, yadav_first-principles_2014}. 

In the [111] direction TiN is built from alternating layers of pure Ti or N; therefore, the (111) surface of a TiN slab is either terminated by a layer of Ti or N. The termination depends on the chemical potential of nitrogen. In an earlier publication~\cite{feldbauer_adhesion_2015} it was shown that both cases are found in reasonable nitrogen concentration ranges by performing a thermodynamic analysis~\cite{wang_hematite_1998, reuter_composition_2001} to calculate the surface Gibbs free energy for the off-stoichiometric slabs~\cite{lee_stoichiometry_2011}. The presented results are in agreement with the Refs.~\cite{liu_first-principles_2004, wang_surface_2010}. In this work the main focus is put on N terminated slabs, since this termination is favorable at small deviations of the nitrogen chemical potential from its molecular reference, i.e. at ambient conditions. Over a wide range of the nitrogen chemical potential the (001) TiN surface is actually the most stable orientation; however, Al/TiN interfaces between (001) slabs do not show any material transfer.~\cite{feldbauer_adhesion_2015} Thus, results for interfaces between (111) surfaces, which show a similar stability at certain values of the chemical potential and allow for material transfer at least between pristine surfaces, are presented in this work. 

Dipole corrections~\cite{neugebauer_adsorbate-substrate_1992} perpendicular to the interface (z direction) were found to be negligible. To decouple the periodically repeated simulation cells in the z direction, a vacuum spacing of at least \unit[10]{\AA} was included in the simulation cell. Performing bulk calculations the lattice parameters of the single materials, \unit[4.040]{\AA} and \unit[4.254]{\AA} for Al and TiN, respectively, were obtained. The experimental lattice constants~\cite{wyckoff_crystal_1971} of Al and TiN are \unit[4.050]{\AA} and \unit[4.265]{\AA}, respectively, confirming our calculation to be accurate within 0.5\% relative error. For the interface simulation cells, an intermediate lattice parameter of \unit[4.144]{\AA} was selected for the lateral xy lattice vectors yielding an equalized relative error of about 2.6\% for both materials. The consequences of such distortions were investigated in detail in an earlier publication~\cite{feldbauer_adhesion_2015}, where no influence of the chosen lattice parameter on the occurrence of material transfer was observed. Assuming a pseudomorphic interface, for the z direction the material-specific values for the layer distances were used. It has to be noted that in reality such a simple mutual adjustment of thick slabs by a combination of stretching and compression does not usually happen, since an incommensurate contact or dislocations at the interface are more plausible. Hence, some of the atoms would not be aligned perfectly according to the stacking of their respective original slabs as well as across the interface, but rather sample slightly different local environments. For computational reasons, these different local arrangements across the interface are assessed by considering various configurations at the interface which constitute limiting cases.

The simulation of the approach and separation of the slabs follows the method described in Ref.~\cite{feldbauer_adhesion_2015}. For the convenience of the reader the main points are repeated here. The approach and subsequent separation of the two slabs were simulated by moving the upper slab in discrete steps along the z direction and allowing for electronic and atomic relaxations after each step. For the atomic relaxations the bottom three Al as well as the top TiN (three Ti and three N) layers were kept frozen at bulk-like distances, while the intermediate ``free'' ones were allowed to fully relax. This is depicted in Fig.~\ref{fig:Al-TiN-initial} for the N-terminated case. For the approach and separation step sizes of \unit[0.2]{\AA} and \unit[0.1]{\AA}, respectively, were used. For the observation of material transfer the separation is more critical; therefore, a smaller step size was used, which allows for higher accuracy but causes increased computational costs. This quasi-static method is used because real systems are considered to behave adiabatically during the investigated processes.~\cite{feldbauer_adhesion_2015} The approach and separation of the slabs were initiated from equilibrium structures, i.e., separately relaxed slabs and the structure with the lowest energy determined during the approach, respectively. To allow for the simulation of a realistic separation process only the topmost, frozen TiN layers were moved in discrete steps in the positive z direction. Again, electronic and atomic relaxations are executed after each step. This procedure was employed to guarantee that a separation of the slabs at the initial interface was not promoted by the enforced step-wise motion. Because of the dimensions of the simulation cell, the transfer of entire atomic layers is investigated. This is seen as a limiting scenario of more realistic interfaces where the transfer of clusters or flakes of atoms is likely. The detailed analysis of such processes, however, goes beyond the scope of this study and would necessarily involve simulations on larger length scales, for example, using classical MD methods. A more detailed discussion on size, strain, and surface effects can be found in an earlier publication.~\cite{feldbauer_adhesion_2015}  

\begin{figure}[hbt]
  \centering
  \includegraphics[width=.2\linewidth]{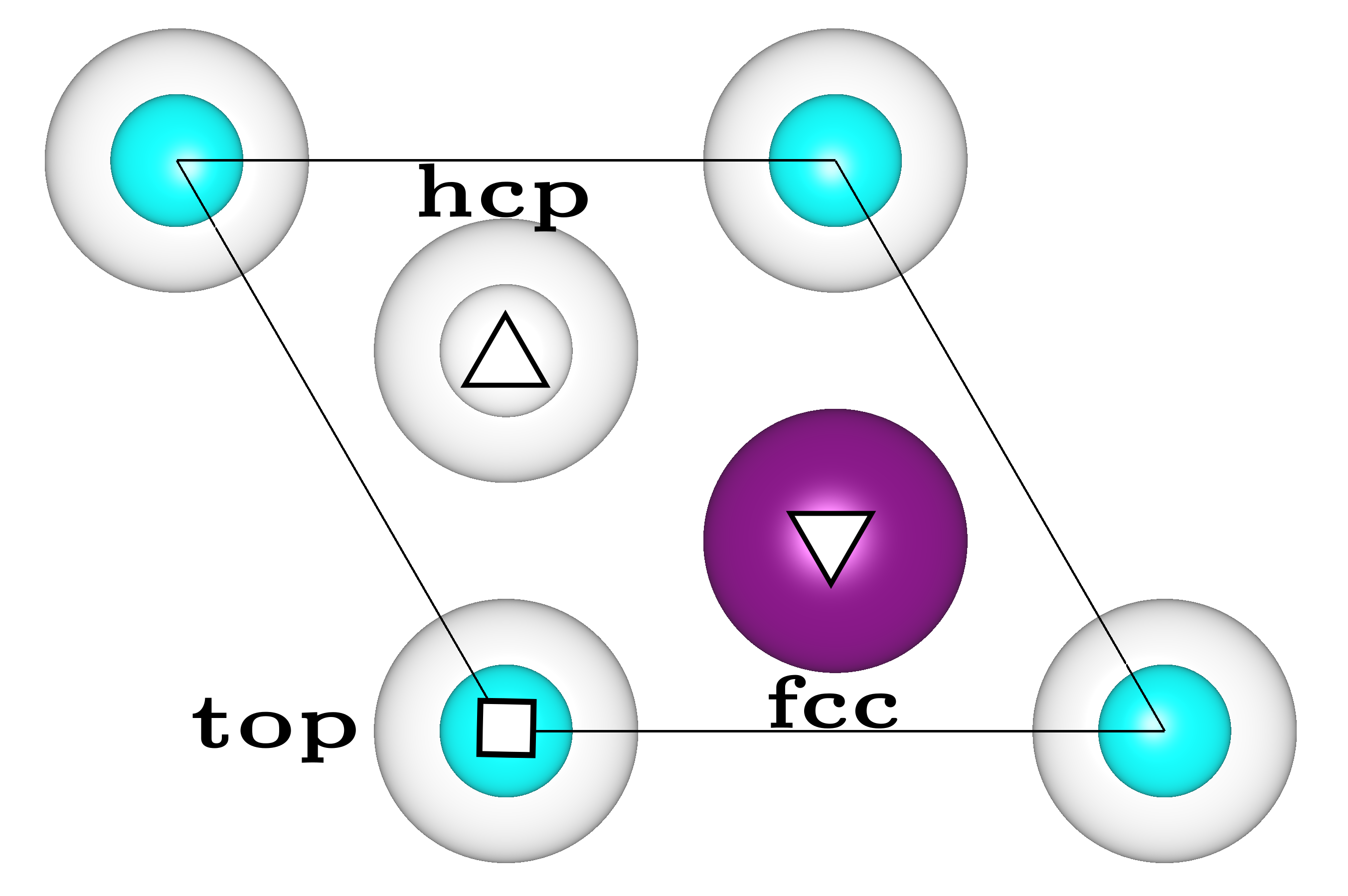}
  \caption{Top view a 1\(\times\)1 surface cell of a N-terminated (111) TiN surface. Filled circles indicate atoms in the top surface layer for each species (Ti and N are given by large purple and small cyan circles, respectively), while empty circles label atoms below the top surface layer. To obtain a Ti-terminated (111) TiN surface the Ti and N atoms of the shown N-terminated surface have to be exchanged. High-symmetry points (top, hcp, fcc) are highlighted.}
  \label{fig:bond-sites}
\end{figure}

To study the effects of different local environments at the interface, several alignments of the slabs were investigated by placing the upper slab laterally  on various sites with respect to the surface of the lower slab. As mentioned above these different configurations represent limiting scenarios for more complex realistic interfaces. The definitions of the alignments are shown in Fig.~\ref{fig:bond-sites} by marking the high-symmetry points on the N-terminated (111) TiN surface where the next Al or O atom can be placed. The alignments are named according to the stacking of the bottom TiN layer with respect to the Al slab, the position of interfacial species is not taken into account in this naming convention. The different configurations are compared via the interaction energy \(E_I(z)\), which is defined as the difference of the total energy of the interacting slabs \(E_{(Al/TiN)}(z)\) at slab height \(z\) and the reference energies  of the two independent slabs, \(E_{(Al)}\) and \(E_{(TiN)}\),
\begin{equation}
  E_I(z) = E_{(Al/TiN)}(z) - E_{(Al)} - E_{(TiN)}.
  \label{equ:interaction-energy}
\end{equation}

\section{Results and Discussion}\label{sec:results}

\subsection{Oxygen adsorption at an Al slab}\label{subsec:oxygen-al}

The fcc adsorption site is favoured by a single oxygen atom on a 1\(\times\)1 (111) Al surface cell. The oxygen is strongly bound to the Al surface exhibiting a rather small binding distance of about \unit[0.7]{\AA} with an adsorption energy of about \unit[-4.7]{eV} with respect to molecular, gas-phase oxygen.~\cite{jacobsen_electronic_1995}  The hcp site is slightly less favourable by about \unit[0.2]{eV}, while the interaction on the top site is much weaker and not favorable considering molecular oxygen as the reference. Additionally, the placement of two O atoms on each 1\(\times\)1 Al surface cell was tested. The two atoms were located initially above the fcc and hcp sites of the Al surface. During a subsequent relaxation the fcc O atom moved beneath the top Al layer, while the hcp O atom stayed above the Al surface. The distance between the top O-Al compound and the remaining O terminated Al slab grew to about \unit[4.2]{\AA}. This large distance combined with a very weak adhesion (\unit[10]{meV}) of this compound on the surface shows that the termination with one O atom per surface cell is preferential on a 1\(\times\)1 Al surface cell. The possibility of an Al$_2$O$_3$-like surface oxide~\cite{jennison_ultrathin_2000} is not considered here. 

Since material transfer at interfaces is of interest, the energy necessary to remove layers from a material is an important quantity.  For this work, additionally the question arises whether this removal energy is affected by an oxygen layer. The removal of the layers was simulated by placing them at a large distance from the remaining slab. The distance was chosen large enough to suppress interactions between the slab and layers. Results for pristine Al slab including the effect of compressive and tensile stress were presented in an earlier publication~\cite{feldbauer_adhesion_2015}. Here, only the values necessary for comparison are repeated in Table~\ref{tab:removal}. Typically, it is energetically unfavourable to remove just one Al layer compared to at least two layers. Moreover, the removal energies are not changing significantly for two or more layers; therefore, the removal of two Al layers is used as a reference here. As mentioned above, oxygen is strongly bound to the Al slab. Thus, it is unlikely that only the oxygen is removed and so the more plausible case of and the removal of the oxygen layer together with Al layers is investigated here. The data in Table~\ref{tab:removal} shows that an oxygen layer hardly influences the removal energies which indicates that the bond breaking between the Al layers is the determining factor. Nevertheless, oxygen may still passivate a surface by suppressing interactions, i.e. weakening the interfacial adhesion, across an interface with another material such as TiN.

\begin{table}[hbt]
  \caption{\label{tab:removal}  Energy costs to remove the top two Al layers, with or without an additional oxygen layer, from a pristine and a oxygenated Al slab, respectively, using PBE and LDA. The removal energies are given in \unit{eV} per 1\(\times\)1 surface cell.}
  \begin{ruledtabular}
    \begin{tabular}{lcc}
      & PBE & LDA \\
      pristine Al & 0.803 & 0.975 \\
      oxygenated Al & 0.804  & 0.988 
    \end{tabular}
  \end{ruledtabular}
\end{table}

\subsection{Lateral Alignments at the Al/TiN Interface}\label{subsec:alignments}

\begin{figure}[hbt]
  \centering
    \includegraphics[width=0.49\linewidth]{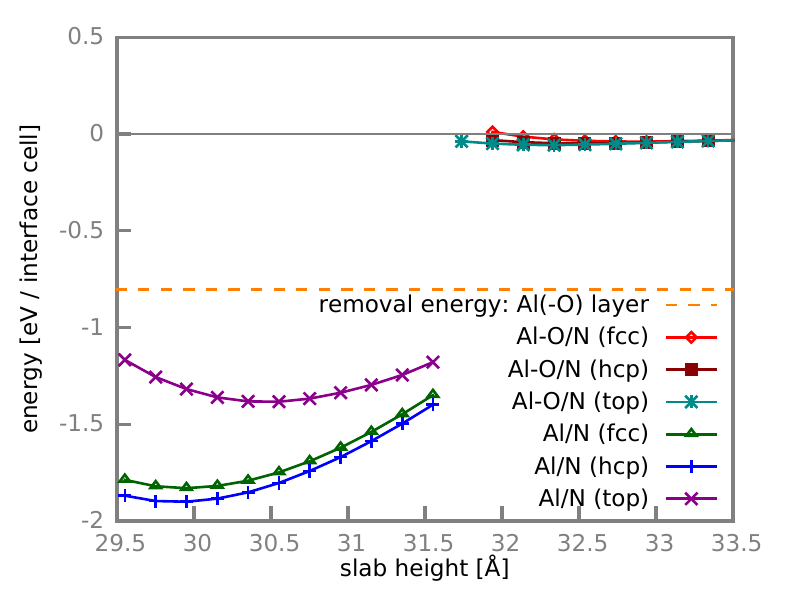}
  \caption{Calculated PBE interaction energies of the relaxed Al/TiN and Al-O/TiN interfaces for the (111) N-terminated surface orientations. Various lateral alignments of the two slabs are considered, see Fig.~\ref{fig:bond-sites}. The horizontal, dashed orange line gives the energy costs to remove at least one Al layer from an (111) Al slab. The Al-O removal energy from an oxygenated Al slab is almost equivalent.}
  \label{fig:bond-sites-pec}
\end{figure}

Effects of different lateral configurations of the slabs at the interface (see Fig.~\ref{fig:bond-sites}) were studied with and without oxygen at the interface. These investigations showed the strong relative dependence of equilibrium properties such as adhesion energies on the chosen alignment. The adhesion energies are energetically one order of magnitude larger for the Al/TiN interface than for the oxygenated contact. The calculated interaction energies (Eq.~\eqref{equ:interaction-energy}), of relaxed Al/TiN and Al-O/TiN interfaces are displayed in Fig.~\ref{fig:bond-sites-pec} for slab heights around the energy minima. For each alignment, these minima are equivalent to the adhesion energies. For the pristine Al slab, in general, the top Al atoms prefer the proximity of N atoms over Ti atoms. More details on interfaces and the bonding situations between pristine Al and TiN slabs can be found in Ref.~\cite{feldbauer_adhesion_2015}. For the oxygenated Al slab in contact with a TiN slab the adhesion energies are strongly reduced compared to the pristine case. The absolute difference between different alignments is, therefore, also rather small. The main focus of this work is on the possibility of material transfer at such interfaces. It has been observed experimentally~\cite{howe_bonding_1993-1, ernst_metal-oxide_1995} as well as in simulations~\cite{feldbauer_adhesion_2015} that metal-ceramic interfaces with weak and strong interfacial adhesion break upon stress at the interface and in bulk areas, respectively. This trend is confirmed by the present calculations. 

As discussed in more detail in reference~\cite{feldbauer_adhesion_2015}, material transfer between the slabs should only occur from an energetical point of view if the energy cost to remove layers is compensated for by the energy gain due to adhesion. This argument is sketched in Fig.~\ref{fig:bond-sites-pec} by including a horizontal line at the negative value of the Al or Al-O removal energy. We find that the pristine and oxygenated surfaces studied here exhibit entirely different behaviour. In the non-oxygenated  case all configurations should lead to the transfer of at least one Al layer, while, the presence of oxygen at the interface should suppress material transfer for all investigated cases.

\begin{figure*}[t]
  \centering
  \begin{subfigure}[b]{0.49\linewidth}
    \includegraphics[width=\linewidth]{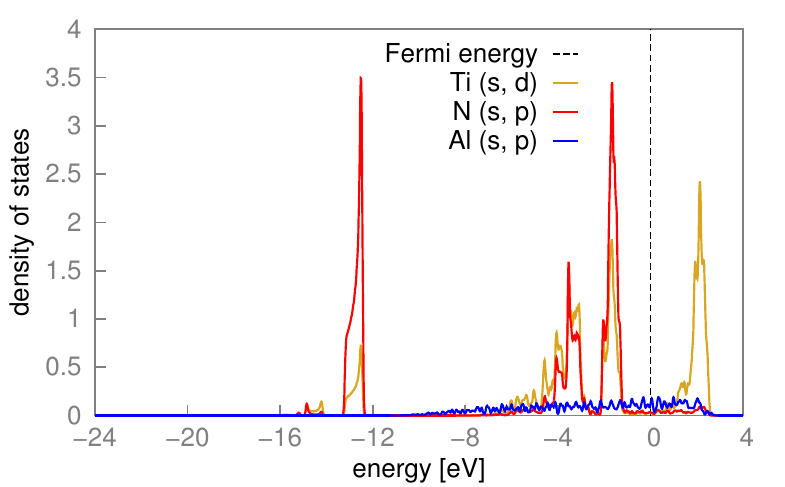}
    \caption{Al \& TiN: surface layers}
    \label{fig:DOS-al-tin-iso-l1}
  \end{subfigure}
  \begin{subfigure}[b]{0.49\linewidth}
    \includegraphics[width=\linewidth]{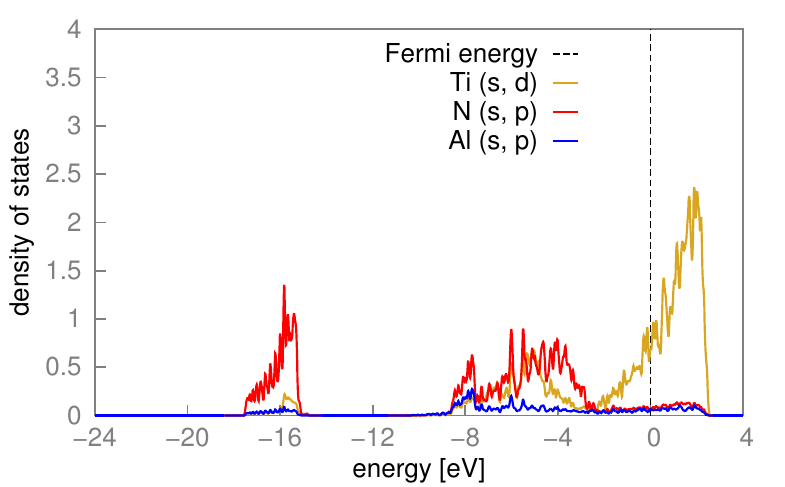}
    \caption{Al/N~(hcp): interface layers}
    \label{fig:DOS-al-h-n-1.4-l1}
  \end{subfigure}
  \caption{Layer-resolved DOSs from PBE calculations for the surface/interface layers of the isolated Al and TiN slabs in (a) as well as of the Al/TiN (111) interface for the Al/N~(hcp) alignment in (b). The Fermi energy is set to \unit[0]{eV}.}
  \label{fig:DOS-al-tin-111}
\end{figure*}

\begin{figure*}[t]
  \centering
  \begin{subfigure}[b]{0.49\linewidth}
    \includegraphics[width=\linewidth]{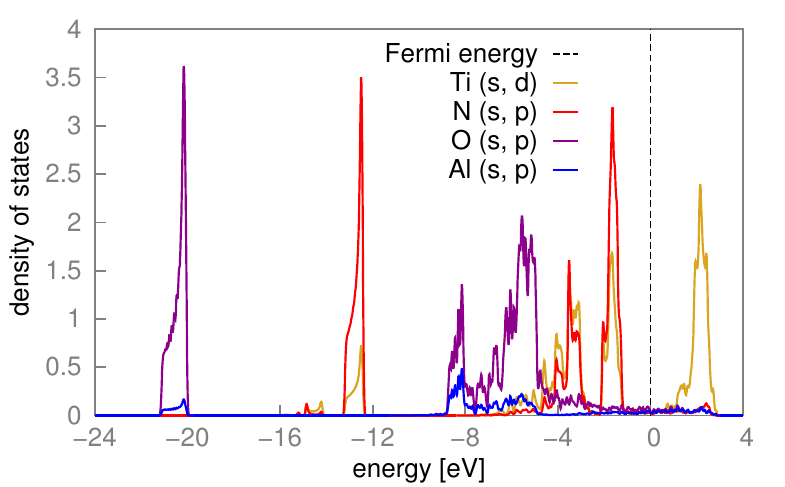}
    \caption{Al-O \& TiN: surface layers}
    \label{fig:DOS-al-o-n-iso-l1}
  \end{subfigure}
  \begin{subfigure}[b]{0.49\linewidth}
    \includegraphics[width=\linewidth]{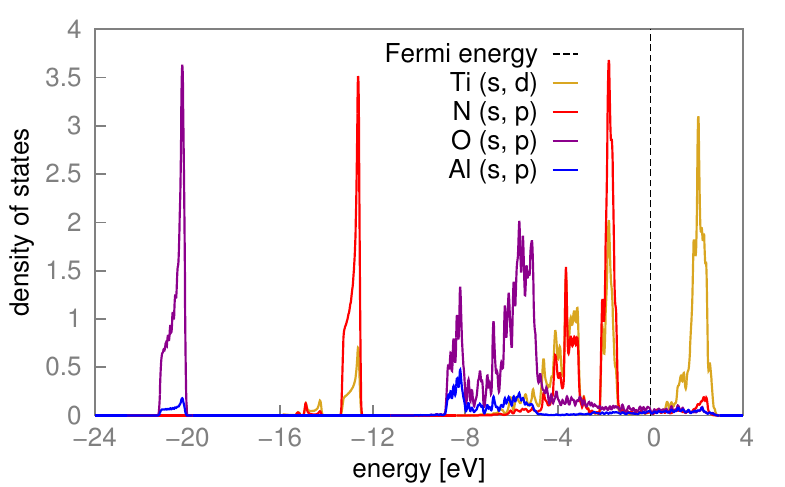}
    \caption{Al-O/N~(top): interface layers}
    \label{fig:DOS-al-o-t-n-3.0-l1}
  \end{subfigure}
  \caption{Layer-resolved DOSs from PBE calculations for the surface/interface layers of the isolated Al-O and TiN slabs in (a) as well as of the Al-O/TiN (111) interface for the Al-O/N~(top) alignments in (b). The Fermi energy is set to \unit[0]{eV}.}
  \label{fig:DOS-al-o-tin-111}
\end{figure*}

For a more comprehensive understanding of the differences between oxygenated and non-oxygenated Al surfaces at Al/TiN interfaces, layer-resolved densities of states (DOSs) and differences in charge densities are investigated. Layer-resolved DOSs of the valence electrons are shown in Fig.~\ref{fig:DOS-al-tin-111} and Fig.~\ref{fig:DOS-al-o-tin-111} for Al/TiN and Al-O/TiN interfaces, respectively. In both figures ``interface (surface) layers'' correspond to the first layers of Al, Ti, and N immediately at the interface (surface). DOSs for the ``sub-interface (sub-surface) layers'', which indicate the next layers of Al, Ti, and N moving deeper into both materials, are given in the supplemental material. Additionally, orbital-resolved DOSs for the surface and interface layers are presented in the supplemental material. For the non-oxygenated Al surface the DOSs of the isolated surfaces and the interface display distinct features, particularly, for the surface and interface layers, see Fig.~\ref{fig:DOS-al-tin-111}. The N sp as well as the Ti sd states are broadened and shifted to lower energies, while the Al sp states show concentrations around \unit[-8]{eV} mainly for the s states as well as \unit[-6]{eV} mainly for the p states and less pronounced around \unit[-16]{eV} resulting in common peaks with N, mainly p, states and Ti, mainly d, states. These changes of the interfacial DOSs indicate a hybridization of Al and N states and explain the strong adhesion because of covalent interactions. The common peaks of Al and Ti states arise because of strong interactions between N and both other atomic species and not due to direct Al-Ti interactions which is in agreement with results for (011) Al/TiN interfaces.~\cite{feldbauer_adhesion_2015} 

For an oxygenated Al surface the consequences of a contact with a TiN surface are quite different, see Fig.~\ref{fig:DOS-al-o-tin-111}. First, the strong interaction between the Al surface and the O monolayer is clearly visible. For comparison, the DOSs of the sub-surface layers can be seen in the supplemental material. The surface layer DOS shows common peaks of the Al sp and O states around \unit[-21]{eV} for O s states as well as \unit[-8]{eV} and \unit[-6]{eV} for O p states explaining the strong bond between Al and O. Furthermore, the decrease of Al states beyond the Fermi energy upon oxygen adsorption indicates that O electrons populate those states. This reduction of available Al states decreases its reactivity. Contrary to the Al/TiN contact, the DOSs of the isolated and interacting Al-O slabs and TiN slabs are almost equivalent and exhibit no indication for a pronounced interaction between an Al-O slab and a TiN slab. This finding agrees well with the very low adhesion energies reported for these interfaces above. The interfacial alignments used here to obtain the DOSs result in the largest adhesion energies for the respective systems. The same trends as described above can be also seen for alignments yielding the weaker interfaces. The corresponding DOSs for these cases are presented in the supplemental material.

\begin{figure}[hbt]
  \centering
  \begin{subfigure}[b]{0.20\linewidth}
    \includegraphics[width=\linewidth]{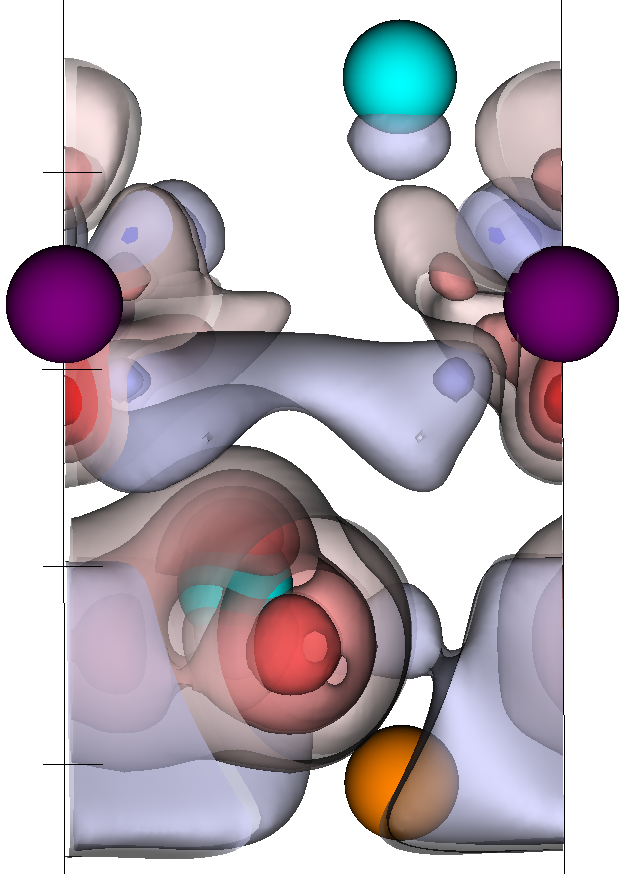}
    \caption{Al/N (hcp)}
    \label{fig:chg-al-n-hcp}
  \end{subfigure}
  \hspace{0.05\linewidth}
  \begin{subfigure}[b]{0.18\linewidth}
    \includegraphics[width=\linewidth]{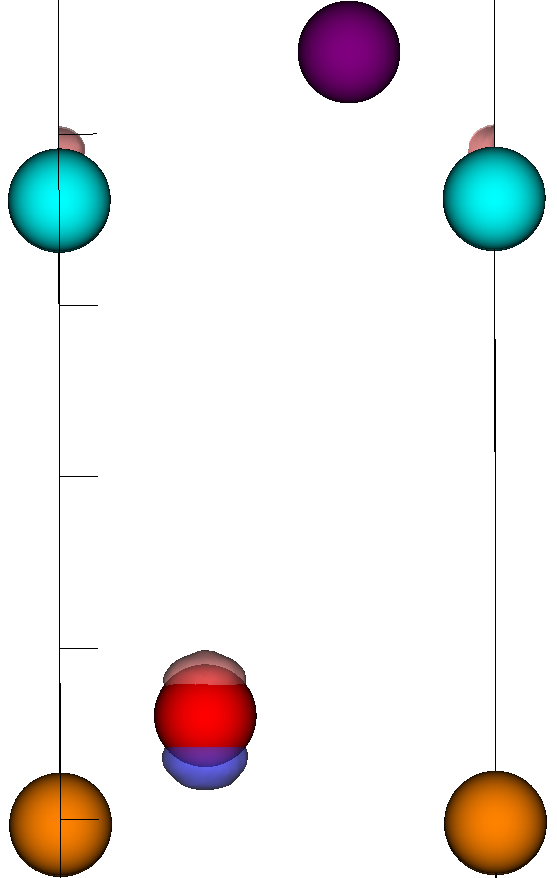}
    \caption{Al-O/N (top)}
    \label{fig:chg-al-o-n-top}
  \end{subfigure}
  \begin{subfigure}[b]{0.4\linewidth}
    \includegraphics[width=\linewidth]{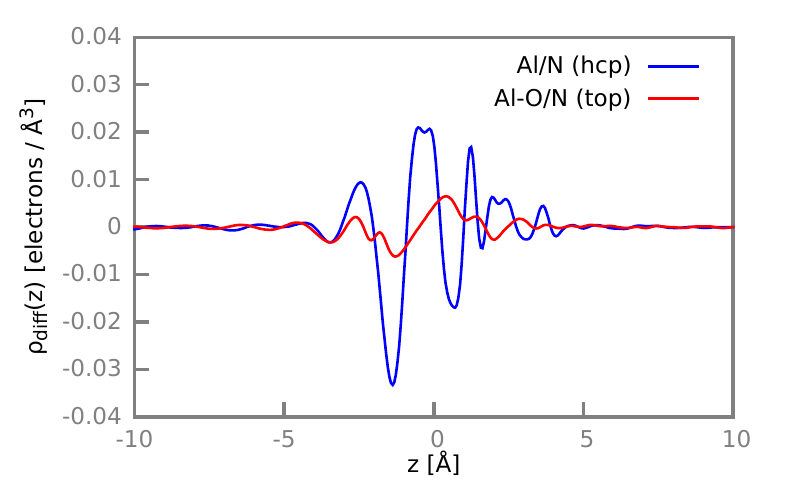}
    \caption{}
    \label{fig:chg-z}
  \end{subfigure}
  \caption{Charge-density differences \(\rho_{diff}\) [see Eq.~\eqref{equ:charge-diff}] at a (111) Al/TiN (N terminated) interface. \(\rho_{diff}\) was obtained from PBE calculations for the relaxed equilibrium configurations of (a) the Al/N~(hcp) alignment and  (b) the Al-O/N~(top) alignment. In (a) and (b), isosurfaces of \(\rho_{diff}\) are plotted for values from -0.2 (blueish, deficit) to 0.3 (reddish, accumulation) electrons/\unit[]{\AA\(^3\)}. The solid black lines indicate the boundaries of the simulation cell. Color code: Al, orange; O, red; Ti, violet; N, cyan. In (c), the profiles of \(\rho_{diff}\) along the [111] direction of both interfaces shown in (a) and (b) are presented. Those profiles  \(\rho_{diff} (z) \) are calculated by two-dimensional integration. Negative values correspond to deficit of electrons. The x-axis gives the vertical distance $z$ (along [111]) from the respective center of each interface.}
  \label{fig:chg-diff-111}
\end{figure}

To support the DOS arguments, charge densities at the interfaces are examined. Particularly, the differences of charge densities \(\rho_{diff}\) between the isolated, independent Al and TiN slabs and the Al/TiN interface are used, which are calculated via
\begin{equation}
  \centering
  \rho_{diff} \,= \, \rho_{Al/TiN} - (\rho_{Al} + \rho_{TiN})\quad,
  \label{equ:charge-diff}
\end{equation}
where \(\rho_{Al/TiN}\) is the charge density of the interface and \(\rho_{Al}\) as well as \(\rho_{TiN}\) are the charge densities of the isolated slabs. Such charge-density differences are visualized as isosurfaces in Fig.~\ref{fig:chg-diff-111} for Al/TiN and Al-O/TiN interfaces. For the non-oxygenated Al surface a significant charge accumulation at an Al/TiN interface is found near the interfacial N atoms pointing towards the neighbouring Al atoms (see Fig.~\ref{fig:chg-al-n-hcp}). This indicates covalent contributions to the interfacial bonding. Furthermore, a charge buildup at the interfacial Ti atoms is observed with contributions parallel to and towards the interface. For clarity, the charge-density differences are integrated in the two dimensions parallel to the interface. The values are normalized to the interfacial cross-section. The results \(\rho_{diff} (z)\) for these integrated charge-density differences along the axis perpendicular to the interface, i.e., in the [111] direction, are shown in Fig.~\ref{fig:chg-z} for a limited region around the interfaces. Such charge-density profiles were used, e.g., to successfully analyse changes in adhesion and corrugation at interfaces~\cite{reguzzoni12}. Both representations in Fig.~\ref{fig:chg-diff-111} indicate a much weaker interaction for oxygenated interfaces. For the oxygenated Al surface the charge-density differences at an Al-O/TiN interface are smaller by about a factor of five; therefore almost no effect is visible in Fig.~\ref{fig:chg-al-o-n-top}. Since Al-O and TiN slabs are used as a reference, the effect of oxygen on the charge distribution in an Al slab is not given. Corroborating the findings from the DOS analysis these charge-density results also suggests a much weaker bond because of the additional oxygen layer.

\subsection{Approach and Separation of Al and TiN Slabs}\label{subsec:interface-loop}

\begin{figure*}[hbt]
  \centering
  \begin{subfigure}[b]{0.45\linewidth}
    \includegraphics[width=\linewidth]{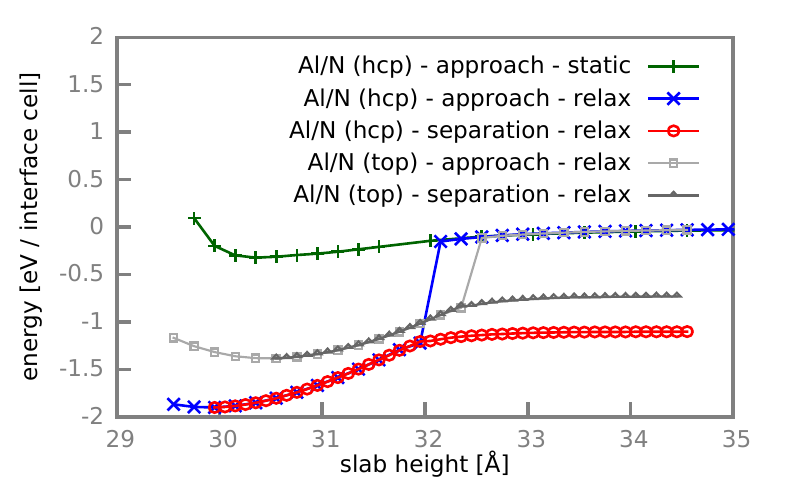}
    \caption{(111) Al/TiN}
    \label{fig:111-al-tin-pec}
  \end{subfigure}
  \begin{subfigure}[b]{0.45\linewidth}
    \includegraphics[width=\linewidth]{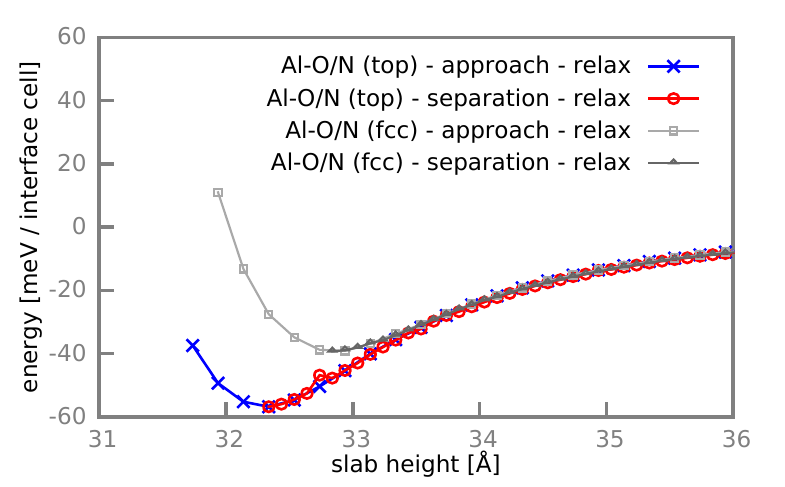}
    \caption{(111) Al-O/TiN}
    \label{fig:111-al-o-tin-pec}
  \end{subfigure}
  \caption{Calculated PBE interaction energies, see Eq.~\eqref{equ:interaction-energy}, for the approach and subsequent separation of (a) Al or (b) Al-O and TiN slabs for (111) N-terminated surface orientations. The alignments follow the definitions in Fig.~\ref{fig:bond-sites}. Mind the different scales on the y-axes in subplots a and b.}
  \label{fig:pecs}
\end{figure*}

Material transfer at interfaces can be studied by ``slowly'', in the sense of using small discrete steps, approaching and subsequently separating the slabs.  Fig.~\ref{fig:pecs} displays the energetical results of such loops for different Al/TiN and Al-O/TiN configurations; the most important information is presented in Table~\ref{tab:loop-data}. The ``approach - static'' curve in Fig.~\ref{fig:pecs} represents a static interaction-energy curve. This data was obtained from calculations where all atoms were kept frozen for each selected interface distance. For large interface distances, differences between static calculations and those including atomic relaxations are small because of the lack of strong interactions between the slabs. For ever-shorter distances, however, the effect of interactions increases, and the relaxed-energy data deviate from the static calculations. This is illustrated by the ``approach - relax'' curves which represent the interaction energies of the approaching slabs including atomic relaxations after each discrete step. Some of these curves exhibit rather large discontinuities which are either due to the Al slab expanding into the gap between the slabs, or due to material transfer between the slabs, particularly, from Al to TiN. Finally, the interaction energies, including atomic relaxations after each step, of the subsequent separation of the slab are given by the ``separation - relax'' curves. These curves feature some kinks mainly because of the breaking apart of the Al/TiN compound into two separated slabs. When material transfer occurs, the separation curves start to deviate from the approach curves. The coloured curves represent configurations with relative strong interfacial interactions, while the grey curves show results for comparably weaker contacts.

\begin{figure}[hbt]
  \centering
  \begin{subfigure}[b]{0.07\linewidth}
    \includegraphics[width=\linewidth]{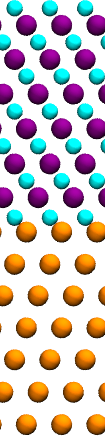}
    \caption{\unit[30.0]{\AA}}
    \label{fig:111-nterm-out1}
  \end{subfigure}
  \hspace{0.02\linewidth}
  \begin{subfigure}[b]{0.07\linewidth}
    \includegraphics[width=\linewidth]{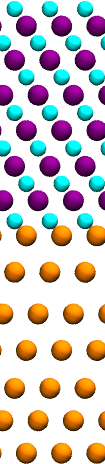}
    \caption{\unit[32.0]{\AA}}
    \label{fig:111-nterm-out2}
  \end{subfigure}
  \hspace{0.02\linewidth}
  \begin{subfigure}[b]{0.07\linewidth}
    \includegraphics[width=\linewidth]{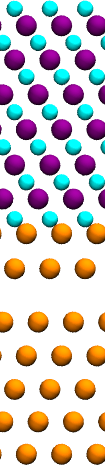}
    \caption{\unit[32.1]{\AA}}
    \label{fig:111-nterm-out3}
  \end{subfigure}
  \hspace{0.02\linewidth}
  \begin{subfigure}[b]{0.07\linewidth}
    \includegraphics[width=\linewidth]{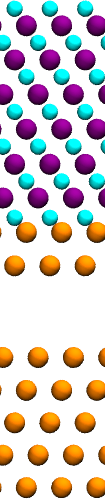}
    \caption{\unit[34.5]{\AA}}
    \label{fig:111-nterm-out4}
  \end{subfigure}
  %\\
  %\vspace{0.5cm}
  \hspace{0.04\linewidth}
  \begin{subfigure}[b]{0.07\linewidth}
    \includegraphics[width=\linewidth]{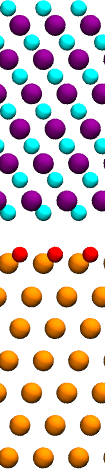}
    \caption{\unit[32.3]{\AA}}
    \label{fig:111-o-nterm-out1}
  \end{subfigure}
  \hspace{0.02\linewidth}
  \begin{subfigure}[b]{0.07\linewidth}
    \includegraphics[width=\linewidth]{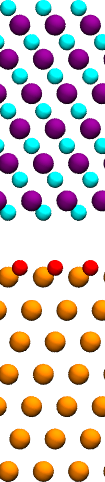}
    \caption{\unit[33.3]{\AA}}
    \label{fig:111-o-nterm-out2}
  \end{subfigure}
  \hspace{0.02\linewidth}
  \begin{subfigure}[b]{0.07\linewidth}
    \includegraphics[width=\linewidth]{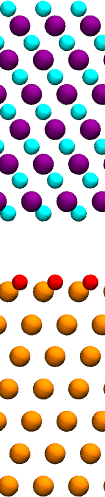}
    \caption{\unit[34.3]{\AA}}
    \label{fig:111-o-nterm-out3}
  \end{subfigure}
  \hspace{0.02\linewidth}
  \begin{subfigure}[b]{0.07\linewidth}
    \includegraphics[width=\linewidth]{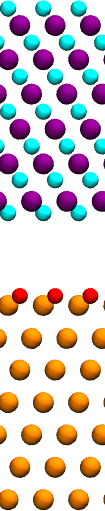}
    \caption{\unit[35.3]{\AA}}
    \label{fig:111-o-nterm-out4}
  \end{subfigure}
  \caption{Separation of Al/N~(hcp) (a--d) and Al-O/N~(top) (e--h)  aligned Al (111) and N-terminated TiN (111) slabs. Al, O, Ti, and N are coloured in orange, red, purple, and cyan, respectively. Subfigures (a) and (e) show the structures at the relaxed equilibrium distance. The steps are defined by the slab height.}
  \label{fig:111-nterm-sep}
\end{figure}

Pristine Al slabs and TiN slabs interact strongly and exhibit adhesion energies of \unit[1--2]{eV} for all alignments at the interface as well as for both TiN terminations.~\cite{feldbauer_adhesion_2015} Since these adhesion energies are always larger than the cost to remove Al layers from an Al slab, material transfer occurs for all configurations upon the separation of the two slabs. Depending on the equilibrium distance between two slabs typically one or two Al layers are transferred, because larger distances hinder the interaction between subinterface layers. Interaction-energy curves are presented in Fig.~\ref{fig:111-al-tin-pec} for the limiting cases represented by the alignments hcp and top which results in the material transfer of two and one Al layer(s), respectively. For the Al/N~(hcp) configuration structural snapshots along the separation are presented in Fig.~\ref{fig:111-nterm-sep} sub-figures (a--d).

The presence of oxygen at the interface between Al and TiN slabs reduces the adhesion energies between Al and TiN slabs significantly. In the case of N-terminated TiN the adhesion energy is diminished to just about \unit[50]{meV} for all investigated alignments at the interface. This value is much smaller than the energy needed to remove material, here O and Al layers, from the oxygenated Al slab which is about \unit[800]{meV}, see Table~\ref{tab:removal}. Thus, no material transfer should occur at these interfaces. Interaction-energy curves displayed in Fig.~\ref{fig:111-al-o-tin-pec} for the approach and subsequent separation of oxygenated Al slabs and TiN slabs exhibit the predicted behaviour indicating no material transfer. Snapshots of the structures during the separation are shown in Fig.~\ref{fig:111-nterm-sep} sub-figures (e--h) for the Al-O/N~(top) configuration. To check consistency, these calculations based on the PBE functional were repeated using LDA. As expected the adhesion energies are increased for all configurations. Moreover, the range of the obtained values from about \unit[100]{meV} to \unit[250]{meV} for the different alignments is larger. The adhesion energies are, however, still significantly lower than the removal energies of about \unit[1]{eV} (see Table~\ref{tab:removal}) and no material transfer occurs. Thus, the LDA results are in agreement with PBE concerning material transfer. In a previous publication, the occurrence of material transfer was demonstrated to be independent of other computational settings, such as the size of the simulation cell as long as the surface structure is preserved.~\cite{feldbauer_adhesion_2015} The obtained results are in line with the findings of other researchers that additional interfacial species like Mg, H, and Zn reduce the adhesion energies at Al/TiN interfaces.~\cite{liu_first-principles_2005, zhang_effects_2007} 

\begin{table}[hbt]
  \caption{\label{tab:loop-data} Equilibrium interface distances d$_\textnormal{0}$ [\unit{\AA}], adhesion energies E$_\textnormal{a}$ [\unit{eV}/interface cell], energy costs to remove layers from the Al slab E$_\textnormal{r}$ [\unit{eV}/interface cell], and number of transferred Al layers (\# TL) for various interface configurations between clean or oxygen covered Al and N-terminated TiN surfaces. Al/Ti and Al/N denote the Ti- and N-terminated surfaces, respectively.}
  \begin{ruledtabular}
    \begin{tabular}{lcccc}
      & d$_\textnormal{0}$  & E$_\textnormal{a}$ & E$_\textnormal{r}$ & \# TL \\
      Al/N~(hcp) & 1.04 & -1.90 & 0.80 &  2 \\
      Al/N~(top) & 1.87 & -1.38 & 0.80 & 1 \\
      Al-O/N~(fcc) & 3.50  & -0.04 & 0.80  &  0   \\
      Al-O/N~(top) & 3.01  & -0.06  & 0.80  & 0    \\
      Al/Ti~(hcp) & 2.22 & -1.78 & 0.80 & 2 \\
      Al/Ti~(top) & 2.67 & -0.94 & 0.80 & 1 \\
      Al-O/Ti~(hcp) & 1.59  & -0.54  & 0.80 & 0    \\
      Al-O/Ti~(top) & 1.61  & -0.61  &  0.80 &  0 
    \end{tabular}
  \end{ruledtabular}
\end{table}

For Ti-terminated TiN slabs the adhesion energies at the Al/TiN interfaces are also reduced upon the introduction of oxygen at the interface. For different alignments at the interface the values of the adhesion energies range between \unit[-0.5]{eV} to \unit[-0.6]{eV} and \unit[-0.9]{eV} to \unit[-1.8]{eV} for the oxygenated and non-oxygenated~\cite{feldbauer_adhesion_2015} contact, respectively. The decrease is less pronounced than for the N-terminated case because of more favourable Ti-O interactions. Nevertheless, material transfer becomes again unfavourable because of the interfacial oxygen, since the adhesion energies become smaller than the Al-O removal energies of about \unit[-0.8]{eV}. The relevant information is summarized in Table~\ref{tab:loop-data}.

\subsection{Cu/C$_{\text{dia}}$ Interfaces}\label{subsec:cu-c}

In addition to the Al/TiN interfaces, Cu/C$_{\text{dia}}$ interfaces were investigated because of their high technological relevance and to test the used method for the simulation of material transfer on a different system. Analogous to the studies of Al/TiN, results on interfaces between the (111) surfaces of Cu/C$_{\text{dia}}$  are discussed in this section. Because of the qualitative similarity of the results and for the sake of space, only some major remarks and results are presented here.

In general, the diamond (111) surface exhibits a (2 $\times$ 1) reconstruction; however, in the case of hydrogen termination as well as at Cu/C$_{\text{dia}}$ interfaces no reconstructions are observed and (1 $\times$ 1) configurations are found~\cite{schaich_structural_1997, wang_copper/diamond_2001}. Thus, in this work only (1 $\times$ 1) structures are considered. At the diamond (111) surface two terminations are possible. Here, the single dangling bond configuration is assumed, because it is more stable for reasonable hydrogen chemical potentials than the triple dangling bond terminations~\cite{wang_copper/diamond_2001, scholze_structure_1996, kern_atomic_1996, kern_atomic_1996-1, kern_atomic_1997}. Particularly, the former configuration is more favourable at interfaces with copper~\cite{wang_copper/diamond_2001}.

The energy necessary to remove material from an (111) Cu slab is about \unit[0.77]{eV}, which will be compared to the adhesion energies for several alignments at Cu/C$_{\text{dia}}$ interfaces in the following. Of the examined high symmetry contact configurations, which are defined in the same way as for the Al/TiN interfaces, the top and fcc alignments result in the strongest and weakest interfaces, respectively. The adhesion energy for the top configuration is about \unit[-1.05]{eV}, while for the fcc case it amounts to about \unit[-0.9]{eV}. Since these adhesion energies are larger than the cost to remove material from the Cu slab, material transfer is expected to occur for all configurations. Employing the method of approach and subsequent separation of the two slabs described above leads indeed to the material transfer of one layer of Cu towards the C$_{\text{dia}}$ slab for all tested contact scenarios, which are considered to be the limiting cases. Figures showing the corresponding interaction-energy curves are included in the supplemental material. These results support the DFT-based work of Guo et al.~\cite{guo_adhesion_2010} who predicted the occurrence of fracture at Cu/C$_{\text{dia}}$ interface between the first two metallic layers or the 2nd and the 3rd layers, because of a bond weakening resulting from the contact with the C$_{\text{dia}}$ counterbody.

As for the Al/TiN interfaces, the effect of additional interfacial species was studied at the contact of Cu and C. Since the dangling bonds at diamond surfaces are often saturated by hydrogen atoms, this species was chosen for our investigation. At the C$_{\text{dia}}$ (111) surface the H atoms prefer to sit at top sites in a distance of just \unit[1.1]{\AA} above each C atom forming a p(1 $\times$ 1) pattern. This is in good agreement with literature, e.g. in Ref.~\cite{wang_copper/diamond_2001}. 

The adhesion energies for all studied contact configurations of the Cu/C$_{\text{dia}}$ interfaces are extremely reduced to about \unit[-0.02]{eV} for hydrogen terminated C$_{\text{dia}}$ surfaces in a similar fashion as for the oxygenated Al surfaces at the Al/TiN interfaces. Since these adhesion energies are much lower than the Cu removal energies of about \unit[0.77]{eV}, no material transfer is expected. This predicted behaviour is verified by the method of approach and separation. Figures of the resulting interaction-energy curves are shown in the supplemental material.

The same results with respect to material transfer were obtained employing a different XC functional, namely LDA. As expected using these functionals the adhesion as well as removal energies for all investigated interfaces are increased similar to the Al/TiN interfaces. For the clean surfaces LDA yields adhesion energies about \unit[-1.21]{eV} and \unit[-1.34]{eV} for the fcc and top alignment, respectively. As for PBE, the adhesion energies are drastically reduced for interfaces involving hydrogenated C$_{\text{dia}}$ surfaces. Using LDA the adhesion energies are between \unit[-0.11]{eV} for the fcc configuration and \unit[-0.13]{eV} for the top case. The energies necessary to remove Cu layers from a Cu slab are also increased to \unit[1.01]{eV} for LDA. These results show that the choice between the XC functionals PBE and LDA does not affect the results for material transfer.

The findings presented above agree well with theoretical and experimental results presented in the introduction including the strong effect of interfacial hydrogen on the adhesion energy~\cite{wang_copper/diamond_2001} and the possibility of a transfer of only one atomic layer of Cu~\cite{zhu_study_2016}. Additionally, performing large-scale molecular dynamics (MD) simulations on the rubbing contact of hydrogenated diamond bodies and tungsten surfaces, Stoyanov et al.~\cite{stoyanov_nanoscale_2014} found that material transfer between C$_{\text{dia}}$ and W mainly occurred at regions where the hydrogen had been depleted from the diamond surface because of the rubbing process. This observation is in qualitative agreement with our study finding material transfer only possible for non-hydrogenated diamond surfaces.

\section{Conclusion and Outlook}\label{sec:conclusion}

The adhesion energies at Al/TiN and Cu/C$_{\text{dia}}$ interfaces are reduced significantly by the presence of a monolayer of oxygen and hydrogen at Al/TiN and Cu/C$_{\text{dia}}$ interfaces, respectively. This reduction is investigated by analysing the densities of states and charge density differences at the aforementioned interfaces which reveal distinct bonding situations. Particularly, the possibility of material transfer at the interfaces is of interest. The occurrence of material transfer was studied by simulating the approach and subsequent separation of two slabs. For the clean surfaces material transfer of one or two atomic layers was found to be favourable for both investigated material combinations depending on the local interfacial configuration. The additional interfacial species O and H passivate the involved surfaces with respect to material transfer because of the strongly reduced adhesion energies for all investigated configurations. This agrees with the observation that metal-ceramic interfaces break in bulk areas or at the interface according to their interfacial adhesion~\cite{howe_bonding_1993-1, ernst_metal-oxide_1995}. The results with respect to material transfer were not affected upon the use of two different approximations for the exchange-correlation functional.

The investigation of more complex interfacial species such as extended oxides instead of the studied monolayers promises to be very interesting in order to model many systems in a more realistic fashion. While, in principle, the method employed in this work can be applied to any pair of materials, in practice this can be computationally quite cumbersome. Large simulation cells, which translate into high computational demands, become necessary for complex materials or materials with an unfavourable bulk lattice mismatch because the translational symmetry has to be preserved and the lattice distortions should be as low as possible. On the other hand, larger cells allow to account for additional features such the minimization of lattice mismatches and the possibility to model dislocations or quasi-incommensurate contacts at interfaces. Furthermore, surface roughness can be treated to some extent by using a stepped surface or a regular arrangement of asperities. 

\section*{Supplementary Data}\label{sec:supp}
  See the Supplementary Data for orbital-resolved {DOSs} of the surface and interface layers of the strong {Al(-O)/TiN} interfaces, for {DOSs} of the weaker {Al(-O)/TiN} interfaces as well as for figures showing interaction-energy curves for the {Cu/C$_{dia}$} interfaces.

\section*{Acknowledgements}\label{sec:acknow}
The authors thank G. Vorlaufer for fruitful discussions. G.F. is grateful for the financial support from the German Research Foundation (DFG) via SFB 986 M$^3$, project A4. G.F., M.W., P.O.B., P.M., and J.R. acknowledge the support by the Austrian Science Fund (FWF): F4109-N28 SFB ViCoM. Part of this work was funded by the Austrian COMET-Program (project K2 XTribology,  No. 849109) via the Austrian Research Promotion Agency (FFG) and  the Province of Nieder\"osterreich, Vorarlberg, and Wien. Part of this work was supported by the European Cooperation in Science and Technology (COST; Action MP1303). The authors also appreciate the ample support of computer resources by the Vienna Scientific Cluster (VSC). Figs.~\ref{fig:Al-TiN-initial} and \ref{fig:bond-sites}  were created employing \textsc{VESTA}~\cite{momma_vesta_2011}, Fig.~\ref{fig:chg-diff-111} utilizing \textsc{VisIt}~\cite{childs_visit:_2012}, and Fig.~\ref{fig:111-nterm-sep} using \textsc{VMD}~\cite{humphrey_vmd:_1996}.

\section*{References}
%\begin{thebibliography}{}%
\bibliographystyle{apsrev4-1}
%\addcontentsline{toc}{section}{References}
\bibliography{refs.bib}

\begin{thebibliography}{96}%
\makeatletter
\providecommand \@ifxundefined [1]{%
 \@ifx{#1\undefined}
}%
\providecommand \@ifnum [1]{%
 \ifnum #1\expandafter \@firstoftwo
 \else \expandafter \@secondoftwo
 \fi
}%
\providecommand \@ifx [1]{%
 \ifx #1\expandafter \@firstoftwo
 \else \expandafter \@secondoftwo
 \fi
}%
\providecommand \natexlab [1]{#1}%
\providecommand \enquote  [1]{``#1''}%
\providecommand \bibnamefont  [1]{#1}%
\providecommand \bibfnamefont [1]{#1}%
\providecommand \citenamefont [1]{#1}%
\providecommand \href@noop [0]{\@secondoftwo}%
\providecommand \href [0]{\begingroup \@sanitize@url \@href}%
\providecommand \@href[1]{\@@startlink{#1}\@@href}%
\providecommand \@@href[1]{\endgroup#1\@@endlink}%
\providecommand \@sanitize@url [0]{\catcode `\\12\catcode `\$12\catcode
  `\&12\catcode `\#12\catcode `\^12\catcode `\_12\catcode `\%12\relax}%
\providecommand \@@startlink[1]{}%
\providecommand \@@endlink[0]{}%
\providecommand \url  [0]{\begingroup\@sanitize@url \@url }%
\providecommand \@url [1]{\endgroup\@href {#1}{\urlprefix }}%
\providecommand \urlprefix  [0]{URL }%
\providecommand \Eprint [0]{\href }%
\providecommand \doibase [0]{http://dx.doi.org/}%
\providecommand \selectlanguage [0]{\@gobble}%
\providecommand \bibinfo  [0]{\@secondoftwo}%
\providecommand \bibfield  [0]{\@secondoftwo}%
\providecommand \translation [1]{[#1]}%
\providecommand \BibitemOpen [0]{}%
\providecommand \bibitemStop [0]{}%
\providecommand \bibitemNoStop [0]{.\EOS\space}%
\providecommand \EOS [0]{\spacefactor3000\relax}%
\providecommand \BibitemShut  [1]{\csname bibitem#1\endcsname}%
\let\auto@bib@innerbib\@empty
%</preamble>
\bibitem [{\citenamefont {Binnig}\ \emph {et~al.}(1986)\citenamefont {Binnig},
  \citenamefont {Quate},\ and\ \citenamefont {Gerber}}]{binnig_atomic_1986}%
  \BibitemOpen
  \bibfield  {author} {\bibinfo {author} {\bibfnamefont {G.}~\bibnamefont
  {Binnig}}, \bibinfo {author} {\bibfnamefont {C.}~\bibnamefont {Quate}}, \
  and\ \bibinfo {author} {\bibfnamefont {C.}~\bibnamefont {Gerber}},\ }\href
  {\doibase 10.1103/PhysRevLett.56.930} {\bibfield  {journal} {\bibinfo
  {journal} {Phys. Rev. Lett.}\ }\textbf {\bibinfo {volume} {56}},\ \bibinfo
  {pages} {930} (\bibinfo {year} {1986})}\BibitemShut {NoStop}%
\bibitem [{\citenamefont {Bennewitz}(2005)}]{bennewitz_friction_2005}%
  \BibitemOpen
  \bibfield  {author} {\bibinfo {author} {\bibfnamefont {R.}~\bibnamefont
  {Bennewitz}},\ }\href {\doibase 10.1016/S1369-7021(05)00845-X} {\bibfield
  {journal} {\bibinfo  {journal} {Mater. Today}\ }\textbf {\bibinfo {volume}
  {8}},\ \bibinfo {pages} {42} (\bibinfo {year} {2005})}\BibitemShut {NoStop}%
\bibitem [{\citenamefont
  {Fischer-Cripps}(2004)}]{fischer-cripps_nanoindentation_2004}%
  \BibitemOpen
  \bibfield  {author} {\bibinfo {author} {\bibfnamefont {A.}~\bibnamefont
  {Fischer-Cripps}},\ }\href {http://www.myilibrary.com?id=954} {\emph
  {\bibinfo {title} {Nanoindentation}}}\ (\bibinfo  {publisher} {Springer},\
  \bibinfo {address} {New York},\ \bibinfo {year} {2004})\BibitemShut {NoStop}%
\bibitem [{\citenamefont {Kim}\ \emph {et~al.}(2007)\citenamefont {Kim},
  \citenamefont {Asay},\ and\ \citenamefont {Dugger}}]{kim_nanotribology_2007}%
  \BibitemOpen
  \bibfield  {author} {\bibinfo {author} {\bibfnamefont {S.}~\bibnamefont
  {Kim}}, \bibinfo {author} {\bibfnamefont {D.}~\bibnamefont {Asay}}, \ and\
  \bibinfo {author} {\bibfnamefont {M.}~\bibnamefont {Dugger}},\ }\href
  {\doibase 10.1016/S1748-0132(07)70140-8} {\bibfield  {journal} {\bibinfo
  {journal} {Nano Today}\ }\textbf {\bibinfo {volume} {2}},\ \bibinfo {pages}
  {22} (\bibinfo {year} {2007})}\BibitemShut {NoStop}%
\bibitem [{\citenamefont {Bhushan}(2008)}]{bhushan_nanotribology_2008}%
  \BibitemOpen
  \bibfield  {author} {\bibinfo {author} {\bibfnamefont {B.}~\bibnamefont
  {Bhushan}},\ }\href {\doibase 10.1098/rsta.2007.2170} {\bibfield  {journal}
  {\bibinfo  {journal} {Philos. Trans. R. Soc., A}\ }\textbf {\bibinfo {volume}
  {366}},\ \bibinfo {pages} {1499} (\bibinfo {year} {2008})}\BibitemShut
  {NoStop}%
\bibitem [{\citenamefont {Bhushan}\ \emph {et~al.}(1995)\citenamefont
  {Bhushan}, \citenamefont {Israelachvili},\ and\ \citenamefont
  {Landman}}]{bhushan_nanotribolgy_1995}%
  \BibitemOpen
  \bibfield  {author} {\bibinfo {author} {\bibfnamefont {B.}~\bibnamefont
  {Bhushan}}, \bibinfo {author} {\bibfnamefont {J.}~\bibnamefont
  {Israelachvili}}, \ and\ \bibinfo {author} {\bibfnamefont {U.}~\bibnamefont
  {Landman}},\ }\href@noop {} {\bibfield  {journal} {\bibinfo  {journal}
  {Nature}\ }\textbf {\bibinfo {volume} {374}},\ \bibinfo {pages} {607}
  (\bibinfo {year} {1995})}\BibitemShut {NoStop}%
\bibitem [{\citenamefont {Gnecco}\ \emph {et~al.}(2002)\citenamefont {Gnecco},
  \citenamefont {Bennewitz},\ and\ \citenamefont
  {Meyer}}]{gnecco_abrasive_2002}%
  \BibitemOpen
  \bibfield  {author} {\bibinfo {author} {\bibfnamefont {E.}~\bibnamefont
  {Gnecco}}, \bibinfo {author} {\bibfnamefont {R.}~\bibnamefont {Bennewitz}}, \
  and\ \bibinfo {author} {\bibfnamefont {E.}~\bibnamefont {Meyer}},\ }\href
  {\doibase 10.1103/PhysRevLett.88.215501} {\bibfield  {journal} {\bibinfo
  {journal} {Phys. Rev. Lett.}\ }\textbf {\bibinfo {volume} {88}},\ \bibinfo
  {pages} {215501} (\bibinfo {year} {2002})}\BibitemShut {NoStop}%
\bibitem [{\citenamefont {Gotsmann}\ and\ \citenamefont
  {Lantz}(2008)}]{gotsmann_atomistic_2008}%
  \BibitemOpen
  \bibfield  {author} {\bibinfo {author} {\bibfnamefont {B.}~\bibnamefont
  {Gotsmann}}\ and\ \bibinfo {author} {\bibfnamefont {M.}~\bibnamefont
  {Lantz}},\ }\href {http://link.aps.org/doi/10.1103/PhysRevLett.101.125501}
  {\bibfield  {journal} {\bibinfo  {journal} {Phys. Rev. Lett.}\ }\textbf
  {\bibinfo {volume} {101}},\ \bibinfo {pages} {125501} (\bibinfo {year}
  {2008})}\BibitemShut {NoStop}%
\bibitem [{\citenamefont {Bhaskaran}\ \emph {et~al.}(2010)\citenamefont
  {Bhaskaran}, \citenamefont {Gotsmann}, \citenamefont {Sebastian},
  \citenamefont {Drechsler}, \citenamefont {Lantz}, \citenamefont {Despont},
  \citenamefont {Jaroenapibal}, \citenamefont {Carpick}, \citenamefont {Chen},\
  and\ \citenamefont {Sridharan}}]{bhaskaran_ultralow_2010}%
  \BibitemOpen
  \bibfield  {author} {\bibinfo {author} {\bibfnamefont {H.}~\bibnamefont
  {Bhaskaran}}, \bibinfo {author} {\bibfnamefont {B.}~\bibnamefont {Gotsmann}},
  \bibinfo {author} {\bibfnamefont {A.}~\bibnamefont {Sebastian}}, \bibinfo
  {author} {\bibfnamefont {U.}~\bibnamefont {Drechsler}}, \bibinfo {author}
  {\bibfnamefont {M.}~\bibnamefont {Lantz}}, \bibinfo {author} {\bibfnamefont
  {M.}~\bibnamefont {Despont}}, \bibinfo {author} {\bibfnamefont
  {P.}~\bibnamefont {Jaroenapibal}}, \bibinfo {author} {\bibfnamefont
  {R.}~\bibnamefont {Carpick}}, \bibinfo {author} {\bibfnamefont
  {Y.}~\bibnamefont {Chen}}, \ and\ \bibinfo {author} {\bibfnamefont
  {K.}~\bibnamefont {Sridharan}},\ }\href {\doibase 10.1038/nnano.2010.3}
  {\bibfield  {journal} {\bibinfo  {journal} {Nat. Nanotechnol.}\ }\textbf
  {\bibinfo {volume} {5}},\ \bibinfo {pages} {181} (\bibinfo {year}
  {2010})}\BibitemShut {NoStop}%
\bibitem [{\citenamefont {Jacobs}\ and\ \citenamefont
  {Carpick}(2013)}]{jacobs_nanoscale_2013}%
  \BibitemOpen
  \bibfield  {author} {\bibinfo {author} {\bibfnamefont {T.}~\bibnamefont
  {Jacobs}}\ and\ \bibinfo {author} {\bibfnamefont {R.}~\bibnamefont
  {Carpick}},\ }\href {\doibase 10.1038/nnano.2012.255} {\bibfield  {journal}
  {\bibinfo  {journal} {Nat. Nanotechnol.}\ }\textbf {\bibinfo {volume} {8}},\
  \bibinfo {pages} {108} (\bibinfo {year} {2013})}\BibitemShut {NoStop}%
\bibitem [{\citenamefont {Kim}\ \emph {et~al.}(2012)\citenamefont {Kim},
  \citenamefont {Yoo},\ and\ \citenamefont {Kim}}]{kim_nano-scale_2012}%
  \BibitemOpen
  \bibfield  {author} {\bibinfo {author} {\bibfnamefont {H.-J.}\ \bibnamefont
  {Kim}}, \bibinfo {author} {\bibfnamefont {S.-S.}\ \bibnamefont {Yoo}}, \ and\
  \bibinfo {author} {\bibfnamefont {D.-E.}\ \bibnamefont {Kim}},\ }\href
  {\doibase 10.1007/s12541-012-0224-y} {\bibfield  {journal} {\bibinfo
  {journal} {International Journal of Precision Engineering and Manufacturing}\
  }\textbf {\bibinfo {volume} {13}},\ \bibinfo {pages} {1709} (\bibinfo {year}
  {2012})}\BibitemShut {NoStop}%
\bibitem [{\citenamefont {Howe}(1993{\natexlab{a}})}]{howe_bonding_1993}%
  \BibitemOpen
  \bibfield  {author} {\bibinfo {author} {\bibfnamefont {J.}~\bibnamefont
  {Howe}},\ }\href@noop {} {\bibfield  {journal} {\bibinfo  {journal} {Int.
  Mater. Rev.}\ }\textbf {\bibinfo {volume} {38}},\ \bibinfo {pages} {233}
  (\bibinfo {year} {1993}{\natexlab{a}})}\BibitemShut {NoStop}%
\bibitem [{\citenamefont {Guo}\ \emph {et~al.}(2010)\citenamefont {Guo},
  \citenamefont {Qi},\ and\ \citenamefont {Li}}]{guo_adhesion_2010}%
  \BibitemOpen
  \bibfield  {author} {\bibinfo {author} {\bibfnamefont {H.}~\bibnamefont
  {Guo}}, \bibinfo {author} {\bibfnamefont {Y.}~\bibnamefont {Qi}}, \ and\
  \bibinfo {author} {\bibfnamefont {X.}~\bibnamefont {Li}},\ }\href {\doibase
  10.1063/1.3277013} {\bibfield  {journal} {\bibinfo  {journal} {J. Appl.
  Phys.}\ }\textbf {\bibinfo {volume} {107}},\ \bibinfo {pages} {033722}
  (\bibinfo {year} {2010})}\BibitemShut {NoStop}%
\bibitem [{\citenamefont {Johansson}(1995)}]{johansson_electronic_1995}%
  \BibitemOpen
  \bibfield  {author} {\bibinfo {author} {\bibfnamefont {L.}~\bibnamefont
  {Johansson}},\ }\href {\doibase 10.1016/0167-5729(94)00005-0} {\bibfield
  {journal} {\bibinfo  {journal} {Surf. Sci. Rep.}\ }\textbf {\bibinfo {volume}
  {21}},\ \bibinfo {pages} {177} (\bibinfo {year} {1995})}\BibitemShut
  {NoStop}%
\bibitem [{\citenamefont {Wang}\ and\ \citenamefont
  {Smith}(2001)}]{wang_copper/diamond_2001}%
  \BibitemOpen
  \bibfield  {author} {\bibinfo {author} {\bibfnamefont {X.-G.}\ \bibnamefont
  {Wang}}\ and\ \bibinfo {author} {\bibfnamefont {J.}~\bibnamefont {Smith}},\
  }\href {\doibase 10.1103/PhysRevLett.87.186103} {\bibfield  {journal}
  {\bibinfo  {journal} {Phys. Rev. Lett.}\ }\textbf {\bibinfo {volume} {87}},\
  \bibinfo {pages} {186103} (\bibinfo {year} {2001})}\BibitemShut {NoStop}%
\bibitem [{\citenamefont {R\"{u}hle}\ \emph {et~al.}(1992)\citenamefont
  {R\"{u}hle}, \citenamefont {Heuer}, \citenamefont {Evans},\ and\
  \citenamefont {Ashby}}]{ruhle_preface_1992}%
  \BibitemOpen
  \bibfield  {author} {\bibinfo {author} {\bibfnamefont {M.}~\bibnamefont
  {R\"{u}hle}}, \bibinfo {author} {\bibfnamefont {A.}~\bibnamefont {Heuer}},
  \bibinfo {author} {\bibfnamefont {A.}~\bibnamefont {Evans}}, \ and\ \bibinfo
  {author} {\bibfnamefont {M.}~\bibnamefont {Ashby}},\ }\href {\doibase
  10.1016/0956-7151(92)90256-E} {\bibfield  {journal} {\bibinfo  {journal}
  {Acta Metall. Mater.}\ }\textbf {\bibinfo {volume} {40}},\ \bibinfo {pages}
  {VII} (\bibinfo {year} {1992})}\BibitemShut {NoStop}%
\bibitem [{\citenamefont {Kawarada}(1996)}]{kawarada_hydrogen-terminated_1996}%
  \BibitemOpen
  \bibfield  {author} {\bibinfo {author} {\bibfnamefont {H.}~\bibnamefont
  {Kawarada}},\ }\href {\doibase 10.1016/S0167-5729(97)80002-7} {\bibfield
  {journal} {\bibinfo  {journal} {Surf. Sci. Rep.}\ }\textbf {\bibinfo {volume}
  {26}},\ \bibinfo {pages} {205} (\bibinfo {year} {1996})}\BibitemShut
  {NoStop}%
\bibitem [{\citenamefont {Vargel}(2004)}]{vargel_corrosion_2004}%
  \BibitemOpen
  \bibfield  {author} {\bibinfo {author} {\bibfnamefont {C.}~\bibnamefont
  {Vargel}},\ }\href {http://www.books24x7.com/marc.asp?bookid=37295} {\emph
  {\bibinfo {title} {Corrosion of Aluminium}}}\ (\bibinfo  {publisher}
  {Elsevier},\ \bibinfo {address} {Amsterdam; Boston},\ \bibinfo {year}
  {2004})\BibitemShut {NoStop}%
\bibitem [{\citenamefont {Thompson}\ and\ \citenamefont
  {Robbins}(1989)}]{thompson_simulations_1989}%
  \BibitemOpen
  \bibfield  {author} {\bibinfo {author} {\bibfnamefont {P.}~\bibnamefont
  {Thompson}}\ and\ \bibinfo {author} {\bibfnamefont {M.}~\bibnamefont
  {Robbins}},\ }\href {\doibase 10.1103/PhysRevLett.63.766} {\bibfield
  {journal} {\bibinfo  {journal} {Phys. Rev. Lett.}\ }\textbf {\bibinfo
  {volume} {63}},\ \bibinfo {pages} {766} (\bibinfo {year} {1989})}\BibitemShut
  {NoStop}%
\bibitem [{\citenamefont {Robbins}\ and\ \citenamefont
  {M\"{u}ser}(2000)}]{bhushan_computer_2000}%
  \BibitemOpen
  \bibfield  {author} {\bibinfo {author} {\bibfnamefont {M.}~\bibnamefont
  {Robbins}}\ and\ \bibinfo {author} {\bibfnamefont {M.}~\bibnamefont
  {M\"{u}ser}},\ }in\ \href
  {http://www.crcnetbase.com/doi/abs/10.1201/9780849377877.ch20} {\emph
  {\bibinfo {booktitle} {{M}odern {T}ribology {H}andbook}}},\ Vol.~\bibinfo
  {volume} {5}\ (\bibinfo  {publisher} {{CRC} Press},\ \bibinfo {year} {2000})\
  pp.\ \bibinfo {pages} {717--765}\BibitemShut {NoStop}%
\bibitem [{\citenamefont {Kenny}\ \emph {et~al.}(2005)\citenamefont {Kenny},
  \citenamefont {Mulliah}, \citenamefont {Sanz-Navarro},\ and\ \citenamefont
  {Smith}}]{kenny_molecular_2005}%
  \BibitemOpen
  \bibfield  {author} {\bibinfo {author} {\bibfnamefont {S.}~\bibnamefont
  {Kenny}}, \bibinfo {author} {\bibfnamefont {D.}~\bibnamefont {Mulliah}},
  \bibinfo {author} {\bibfnamefont {C.}~\bibnamefont {Sanz-Navarro}}, \ and\
  \bibinfo {author} {\bibfnamefont {R.}~\bibnamefont {Smith}},\ }\href
  {\doibase 10.1098/rsta.2005.1621} {\bibfield  {journal} {\bibinfo  {journal}
  {Philos. Trans. R. Soc., A}\ }\textbf {\bibinfo {volume} {363}},\ \bibinfo
  {pages} {1949} (\bibinfo {year} {2005})}\BibitemShut {NoStop}%
\bibitem [{\citenamefont {Schall}\ \emph {et~al.}(2007)\citenamefont {Schall},
  \citenamefont {Mikulski}, \citenamefont {Chateauneuf}, \citenamefont {Gao},\
  and\ \citenamefont {Harrison}}]{david_schall_molecular_2007}%
  \BibitemOpen
  \bibfield  {author} {\bibinfo {author} {\bibfnamefont {J.}~\bibnamefont
  {Schall}}, \bibinfo {author} {\bibfnamefont {P.}~\bibnamefont {Mikulski}},
  \bibinfo {author} {\bibfnamefont {G.}~\bibnamefont {Chateauneuf}}, \bibinfo
  {author} {\bibfnamefont {G.}~\bibnamefont {Gao}}, \ and\ \bibinfo {author}
  {\bibfnamefont {J.}~\bibnamefont {Harrison}},\ }in\ \href
  {http://linkinghub.elsevier.com/retrieve/pii/B9780444527721500366} {\emph
  {\bibinfo {booktitle} {Superlubricity}}}\ (\bibinfo  {publisher} {Elsevier},\
  \bibinfo {year} {2007})\ pp.\ \bibinfo {pages} {79--102}\BibitemShut
  {NoStop}%
\bibitem [{\citenamefont {Szlufarska}\ \emph {et~al.}(2008)\citenamefont
  {Szlufarska}, \citenamefont {Chandross},\ and\ \citenamefont
  {Carpick}}]{szlufarska_recent_2008}%
  \BibitemOpen
  \bibfield  {author} {\bibinfo {author} {\bibfnamefont {I.}~\bibnamefont
  {Szlufarska}}, \bibinfo {author} {\bibfnamefont {M.}~\bibnamefont
  {Chandross}}, \ and\ \bibinfo {author} {\bibfnamefont {R.}~\bibnamefont
  {Carpick}},\ }\href {\doibase 10.1088/0022-3727/41/12/123001} {\bibfield
  {journal} {\bibinfo  {journal} {J. Phys. D: Appl. Phys.}\ }\textbf {\bibinfo
  {volume} {41}},\ \bibinfo {pages} {123001} (\bibinfo {year}
  {2008})}\BibitemShut {NoStop}%
\bibitem [{\citenamefont {Vernes}\ \emph {et~al.}(2012)\citenamefont {Vernes},
  \citenamefont {Eder}, \citenamefont {Vorlaufer},\ and\ \citenamefont
  {Betz}}]{vernes_three-term_2012}%
  \BibitemOpen
  \bibfield  {author} {\bibinfo {author} {\bibfnamefont {A.}~\bibnamefont
  {Vernes}}, \bibinfo {author} {\bibfnamefont {S.}~\bibnamefont {Eder}},
  \bibinfo {author} {\bibfnamefont {G.}~\bibnamefont {Vorlaufer}}, \ and\
  \bibinfo {author} {\bibfnamefont {G.}~\bibnamefont {Betz}},\ }\href {\doibase
  10.1039/c2fd00120a} {\bibfield  {journal} {\bibinfo  {journal} {Faraday
  Discuss.}\ }\textbf {\bibinfo {volume} {156}},\ \bibinfo {pages} {173}
  (\bibinfo {year} {2012})}\BibitemShut {NoStop}%
\bibitem [{\citenamefont {Eder}\ \emph
  {et~al.}(2015{\natexlab{a}})\citenamefont {Eder}, \citenamefont {Feldbauer},
  \citenamefont {Bianchi}, \citenamefont {Cihak-Bayr}, \citenamefont {Betz},\
  and\ \citenamefont {Vernes}}]{eder_applicability_2015}%
  \BibitemOpen
  \bibfield  {author} {\bibinfo {author} {\bibfnamefont {S.}~\bibnamefont
  {Eder}}, \bibinfo {author} {\bibfnamefont {G.}~\bibnamefont {Feldbauer}},
  \bibinfo {author} {\bibfnamefont {D.}~\bibnamefont {Bianchi}}, \bibinfo
  {author} {\bibfnamefont {U.}~\bibnamefont {Cihak-Bayr}}, \bibinfo {author}
  {\bibfnamefont {G.}~\bibnamefont {Betz}}, \ and\ \bibinfo {author}
  {\bibfnamefont {A.}~\bibnamefont {Vernes}},\ }\href {\doibase
  10.1103/PhysRevLett.115.025502} {\bibfield  {journal} {\bibinfo  {journal}
  {Phys. Rev. Lett.}\ }\textbf {\bibinfo {volume} {115}},\ \bibinfo {pages}
  {025502} (\bibinfo {year} {2015}{\natexlab{a}})}\BibitemShut {NoStop}%
\bibitem [{\citenamefont {Eder}\ \emph
  {et~al.}(2015{\natexlab{b}})\citenamefont {Eder}, \citenamefont {Cihak-Bayr},
  \citenamefont {Vernes},\ and\ \citenamefont {Betz}}]{eder_evolution_2015}%
  \BibitemOpen
  \bibfield  {author} {\bibinfo {author} {\bibfnamefont {S.}~\bibnamefont
  {Eder}}, \bibinfo {author} {\bibfnamefont {U.}~\bibnamefont {Cihak-Bayr}},
  \bibinfo {author} {\bibfnamefont {A.}~\bibnamefont {Vernes}}, \ and\ \bibinfo
  {author} {\bibfnamefont {G.}~\bibnamefont {Betz}},\ }\href {\doibase
  10.1088/0022-3727/48/46/465308} {\bibfield  {journal} {\bibinfo  {journal}
  {J. Phys. D: Appl. Phys.}\ }\textbf {\bibinfo {volume} {48}},\ \bibinfo
  {pages} {465308} (\bibinfo {year} {2015}{\natexlab{b}})}\BibitemShut
  {NoStop}%
\bibitem [{\citenamefont {Eder}\ \emph {et~al.}(2017)\citenamefont {Eder},
  \citenamefont {Bianchi}, \citenamefont {Cihak-Bayr},\ and\ \citenamefont
  {Gkagkas}}]{eder_methods_2016}%
  \BibitemOpen
  \bibfield  {author} {\bibinfo {author} {\bibfnamefont {S.}~\bibnamefont
  {Eder}}, \bibinfo {author} {\bibfnamefont {D.}~\bibnamefont {Bianchi}},
  \bibinfo {author} {\bibfnamefont {U.}~\bibnamefont {Cihak-Bayr}}, \ and\
  \bibinfo {author} {\bibfnamefont {K.}~\bibnamefont {Gkagkas}},\ }\href
  {\doibase 10.1016/j.cpc.2016.10.017} {\bibfield  {journal} {\bibinfo
  {journal} {Comput. Phys. Commun.}\ }\textbf {\bibinfo {volume} {212}},\
  \bibinfo {pages} {100} (\bibinfo {year} {2017})}\BibitemShut {NoStop}%
\bibitem [{\citenamefont {Zhong}\ and\ \citenamefont
  {Tom\'{a}nek}(1990)}]{zhong_first-principles_1990}%
  \BibitemOpen
  \bibfield  {author} {\bibinfo {author} {\bibfnamefont {W.}~\bibnamefont
  {Zhong}}\ and\ \bibinfo {author} {\bibfnamefont {D.}~\bibnamefont
  {Tom\'{a}nek}},\ }\href {\doibase 10.1103/PhysRevLett.64.3054} {\bibfield
  {journal} {\bibinfo  {journal} {Phys. Rev. Lett.}\ }\textbf {\bibinfo
  {volume} {64}},\ \bibinfo {pages} {3054} (\bibinfo {year}
  {1990})}\BibitemShut {NoStop}%
\bibitem [{\citenamefont {Dag}\ and\ \citenamefont
  {Ciraci}(2004)}]{dag_atomic_2004}%
  \BibitemOpen
  \bibfield  {author} {\bibinfo {author} {\bibfnamefont {S.}~\bibnamefont
  {Dag}}\ and\ \bibinfo {author} {\bibfnamefont {S.}~\bibnamefont {Ciraci}},\
  }\href {\doibase 10.1103/PhysRevB.70.241401} {\bibfield  {journal} {\bibinfo
  {journal} {Phys. Rev. B}\ }\textbf {\bibinfo {volume} {70}},\ \bibinfo
  {pages} {241401} (\bibinfo {year} {2004})}\BibitemShut {NoStop}%
\bibitem [{\citenamefont {Garvey}\ \emph {et~al.}(2011)\citenamefont {Garvey},
  \citenamefont {Furlong}, \citenamefont {Weinert},\ and\ \citenamefont
  {Tysoe}}]{garvey_shear_2011}%
  \BibitemOpen
  \bibfield  {author} {\bibinfo {author} {\bibfnamefont {M.}~\bibnamefont
  {Garvey}}, \bibinfo {author} {\bibfnamefont {O.}~\bibnamefont {Furlong}},
  \bibinfo {author} {\bibfnamefont {M.}~\bibnamefont {Weinert}}, \ and\
  \bibinfo {author} {\bibfnamefont {W.}~\bibnamefont {Tysoe}},\ }\href
  {\doibase 10.1088/0953-8984/23/26/265003} {\bibfield  {journal} {\bibinfo
  {journal} {J. Phys.: Condens. Matter}\ }\textbf {\bibinfo {volume} {23}},\
  \bibinfo {pages} {265003} (\bibinfo {year} {2011})}\BibitemShut {NoStop}%
\bibitem [{\citenamefont {Zilibotti}\ and\ \citenamefont
  {Righi}(2011)}]{zilibotti_ab_2011}%
  \BibitemOpen
  \bibfield  {author} {\bibinfo {author} {\bibfnamefont {G.}~\bibnamefont
  {Zilibotti}}\ and\ \bibinfo {author} {\bibfnamefont {M.}~\bibnamefont
  {Righi}},\ }\href {\doibase 10.1021/la200783a} {\bibfield  {journal}
  {\bibinfo  {journal} {Langmuir}\ }\textbf {\bibinfo {volume} {27}},\ \bibinfo
  {pages} {6862} (\bibinfo {year} {2011})}\BibitemShut {NoStop}%
\bibitem [{\citenamefont {Cahangirov}\ \emph {et~al.}(2012)\citenamefont
  {Cahangirov}, \citenamefont {Ataca}, \citenamefont {Topsakal}, \citenamefont
  {Sahin},\ and\ \citenamefont {Ciraci}}]{cahangirov_frictional_2012}%
  \BibitemOpen
  \bibfield  {author} {\bibinfo {author} {\bibfnamefont {S.}~\bibnamefont
  {Cahangirov}}, \bibinfo {author} {\bibfnamefont {C.}~\bibnamefont {Ataca}},
  \bibinfo {author} {\bibfnamefont {M.}~\bibnamefont {Topsakal}}, \bibinfo
  {author} {\bibfnamefont {H.}~\bibnamefont {Sahin}}, \ and\ \bibinfo {author}
  {\bibfnamefont {S.}~\bibnamefont {Ciraci}},\ }\href {\doibase
  10.1103/PhysRevLett.108.126103} {\bibfield  {journal} {\bibinfo  {journal}
  {Phys. Rev. Lett.}\ }\textbf {\bibinfo {volume} {108}},\ \bibinfo {pages}
  {126103} (\bibinfo {year} {2012})}\BibitemShut {NoStop}%
\bibitem [{\citenamefont {Kwon}\ \emph {et~al.}(2012)\citenamefont {Kwon},
  \citenamefont {Ko}, \citenamefont {Jeon}, \citenamefont {Kim},\ and\
  \citenamefont {Park}}]{kwon_enhanced_2012}%
  \BibitemOpen
  \bibfield  {author} {\bibinfo {author} {\bibfnamefont {S.}~\bibnamefont
  {Kwon}}, \bibinfo {author} {\bibfnamefont {J.-H.}\ \bibnamefont {Ko}},
  \bibinfo {author} {\bibfnamefont {K.-J.}\ \bibnamefont {Jeon}}, \bibinfo
  {author} {\bibfnamefont {Y.-H.}\ \bibnamefont {Kim}}, \ and\ \bibinfo
  {author} {\bibfnamefont {J.}~\bibnamefont {Park}},\ }\href {\doibase
  10.1021/nl204019k} {\bibfield  {journal} {\bibinfo  {journal} {Nano Letters}\
  }\textbf {\bibinfo {volume} {12}},\ \bibinfo {pages} {6043} (\bibinfo {year}
  {2012})}\BibitemShut {NoStop}%
\bibitem [{\citenamefont {Bedolla}\ \emph {et~al.}(2014)\citenamefont
  {Bedolla}, \citenamefont {Feldbauer}, \citenamefont {Wolloch}, \citenamefont
  {Gruber}, \citenamefont {Eder}, \citenamefont {D\"{o}rr}, \citenamefont
  {Mohn}, \citenamefont {Redinger},\ and\ \citenamefont
  {Vernes}}]{bedolla_density_2014}%
  \BibitemOpen
  \bibfield  {author} {\bibinfo {author} {\bibfnamefont {P.}~\bibnamefont
  {Bedolla}}, \bibinfo {author} {\bibfnamefont {G.}~\bibnamefont {Feldbauer}},
  \bibinfo {author} {\bibfnamefont {M.}~\bibnamefont {Wolloch}}, \bibinfo
  {author} {\bibfnamefont {C.}~\bibnamefont {Gruber}}, \bibinfo {author}
  {\bibfnamefont {S.}~\bibnamefont {Eder}}, \bibinfo {author} {\bibfnamefont
  {N.}~\bibnamefont {D\"{o}rr}}, \bibinfo {author} {\bibfnamefont
  {P.}~\bibnamefont {Mohn}}, \bibinfo {author} {\bibfnamefont {J.}~\bibnamefont
  {Redinger}}, \ and\ \bibinfo {author} {\bibfnamefont {A.}~\bibnamefont
  {Vernes}},\ }\href {\doibase 10.1021/jp504695m} {\bibfield  {journal}
  {\bibinfo  {journal} {J. Phys. Chem. C}\ }\textbf {\bibinfo {volume} {118}},\
  \bibinfo {pages} {21428} (\bibinfo {year} {2014})}\BibitemShut {NoStop}%
\bibitem [{\citenamefont {Wolloch}\ \emph {et~al.}(2014)\citenamefont
  {Wolloch}, \citenamefont {Feldbauer}, \citenamefont {Mohn}, \citenamefont
  {Redinger},\ and\ \citenamefont {Vernes}}]{wolloch_ab-initio_2014}%
  \BibitemOpen
  \bibfield  {author} {\bibinfo {author} {\bibfnamefont {M.}~\bibnamefont
  {Wolloch}}, \bibinfo {author} {\bibfnamefont {G.}~\bibnamefont {Feldbauer}},
  \bibinfo {author} {\bibfnamefont {P.}~\bibnamefont {Mohn}}, \bibinfo {author}
  {\bibfnamefont {J.}~\bibnamefont {Redinger}}, \ and\ \bibinfo {author}
  {\bibfnamefont {A.}~\bibnamefont {Vernes}},\ }\href {\doibase
  10.1103/PhysRevB.90.195418} {\bibfield  {journal} {\bibinfo  {journal} {Phys.
  Rev. B}\ }\textbf {\bibinfo {volume} {90}},\ \bibinfo {pages} {195418}
  (\bibinfo {year} {2014})}\BibitemShut {NoStop}%
\bibitem [{\citenamefont {Feldbauer}\ \emph {et~al.}(2015)\citenamefont
  {Feldbauer}, \citenamefont {Wolloch}, \citenamefont {Bedolla}, \citenamefont
  {Mohn}, \citenamefont {Redinger},\ and\ \citenamefont
  {Vernes}}]{feldbauer_adhesion_2015}%
  \BibitemOpen
  \bibfield  {author} {\bibinfo {author} {\bibfnamefont {G.}~\bibnamefont
  {Feldbauer}}, \bibinfo {author} {\bibfnamefont {M.}~\bibnamefont {Wolloch}},
  \bibinfo {author} {\bibfnamefont {P.}~\bibnamefont {Bedolla}}, \bibinfo
  {author} {\bibfnamefont {P.}~\bibnamefont {Mohn}}, \bibinfo {author}
  {\bibfnamefont {J.}~\bibnamefont {Redinger}}, \ and\ \bibinfo {author}
  {\bibfnamefont {A.}~\bibnamefont {Vernes}},\ }\href@noop {} {\bibfield
  {journal} {\bibinfo  {journal} {Phys. Rev. B}\ }\textbf {\bibinfo {volume}
  {91}},\ \bibinfo {pages} {165413} (\bibinfo {year} {2015})}\BibitemShut
  {NoStop}%
\bibitem [{\citenamefont {Wolloch}\ \emph {et~al.}(2015)\citenamefont
  {Wolloch}, \citenamefont {Feldbauer}, \citenamefont {Mohn}, \citenamefont
  {Redinger},\ and\ \citenamefont {Vernes}}]{wolloch_ab_2015}%
  \BibitemOpen
  \bibfield  {author} {\bibinfo {author} {\bibfnamefont {M.}~\bibnamefont
  {Wolloch}}, \bibinfo {author} {\bibfnamefont {G.}~\bibnamefont {Feldbauer}},
  \bibinfo {author} {\bibfnamefont {P.}~\bibnamefont {Mohn}}, \bibinfo {author}
  {\bibfnamefont {J.}~\bibnamefont {Redinger}}, \ and\ \bibinfo {author}
  {\bibfnamefont {A.}~\bibnamefont {Vernes}},\ }\href {\doibase
  10.1103/PhysRevB.91.195436} {\bibfield  {journal} {\bibinfo  {journal} {Phys.
  Rev. B}\ }\textbf {\bibinfo {volume} {91}},\ \bibinfo {pages} {195436}
  (\bibinfo {year} {2015})}\BibitemShut {NoStop}%
\bibitem [{\citenamefont {Finnis}(1996)}]{finnis_theory_1996}%
  \BibitemOpen
  \bibfield  {author} {\bibinfo {author} {\bibfnamefont {M.}~\bibnamefont
  {Finnis}},\ }\href {\doibase 10.1088/0953-8984/8/32/003} {\bibfield
  {journal} {\bibinfo  {journal} {J. Phys.: Condens. Matter}\ }\textbf
  {\bibinfo {volume} {8}},\ \bibinfo {pages} {5811} (\bibinfo {year}
  {1996})}\BibitemShut {NoStop}%
\bibitem [{\citenamefont {Lundqvist}\ \emph {et~al.}(2001)\citenamefont
  {Lundqvist}, \citenamefont {Bogicevic}, \citenamefont {Carling},
  \citenamefont {Dudiy}, \citenamefont {Gao}, \citenamefont {Hartford},
  \citenamefont {Hyldgaard}, \citenamefont {Jacobson}, \citenamefont
  {Langreth}, \citenamefont {Lorente}, \citenamefont {Ovesson}, \citenamefont
  {Razaznejad}, \citenamefont {Ruberto}, \citenamefont {Rydberg}, \citenamefont
  {Schr\"{o}der}, \citenamefont {Simak}, \citenamefont {Wahnstr\"{o}m},\ and\
  \citenamefont {Yourdshahyan}}]{lundqvist_density-functional_2001}%
  \BibitemOpen
  \bibfield  {author} {\bibinfo {author} {\bibfnamefont {B.}~\bibnamefont
  {Lundqvist}}, \bibinfo {author} {\bibfnamefont {A.}~\bibnamefont
  {Bogicevic}}, \bibinfo {author} {\bibfnamefont {K.}~\bibnamefont {Carling}},
  \bibinfo {author} {\bibfnamefont {S.}~\bibnamefont {Dudiy}}, \bibinfo
  {author} {\bibfnamefont {S.}~\bibnamefont {Gao}}, \bibinfo {author}
  {\bibfnamefont {J.}~\bibnamefont {Hartford}}, \bibinfo {author}
  {\bibfnamefont {P.}~\bibnamefont {Hyldgaard}}, \bibinfo {author}
  {\bibfnamefont {N.}~\bibnamefont {Jacobson}}, \bibinfo {author}
  {\bibfnamefont {D.}~\bibnamefont {Langreth}}, \bibinfo {author}
  {\bibfnamefont {N.}~\bibnamefont {Lorente}}, \bibinfo {author} {\bibfnamefont
  {S.}~\bibnamefont {Ovesson}}, \bibinfo {author} {\bibfnamefont
  {B.}~\bibnamefont {Razaznejad}}, \bibinfo {author} {\bibfnamefont
  {C.}~\bibnamefont {Ruberto}}, \bibinfo {author} {\bibfnamefont
  {H.}~\bibnamefont {Rydberg}}, \bibinfo {author} {\bibfnamefont
  {E.}~\bibnamefont {Schr\"{o}der}}, \bibinfo {author} {\bibfnamefont
  {S.}~\bibnamefont {Simak}}, \bibinfo {author} {\bibfnamefont
  {G.}~\bibnamefont {Wahnstr\"{o}m}}, \ and\ \bibinfo {author} {\bibfnamefont
  {Y.}~\bibnamefont {Yourdshahyan}},\ }\href {\doibase
  10.1016/S0039-6028(01)01225-0} {\bibfield  {journal} {\bibinfo  {journal}
  {Surf. Sci.}\ }\textbf {\bibinfo {volume} {493}},\ \bibinfo {pages} {253}
  (\bibinfo {year} {2001})}\BibitemShut {NoStop}%
\bibitem [{\citenamefont {Sinnott}\ and\ \citenamefont
  {Dickey}(2003)}]{sinnott_ceramic/metal_2003}%
  \BibitemOpen
  \bibfield  {author} {\bibinfo {author} {\bibfnamefont {S.}~\bibnamefont
  {Sinnott}}\ and\ \bibinfo {author} {\bibfnamefont {E.}~\bibnamefont
  {Dickey}},\ }\href {\doibase 10.1016/j.mser.2003.09.001} {\bibfield
  {journal} {\bibinfo  {journal} {Mater. Sci. Eng., R}\ }\textbf {\bibinfo
  {volume} {43}},\ \bibinfo {pages} {1} (\bibinfo {year} {2003})}\BibitemShut
  {NoStop}%
\bibitem [{\citenamefont {Ercolessi}\ and\ \citenamefont
  {Adams}(1994)}]{ercolessi_interatomic_1994}%
  \BibitemOpen
  \bibfield  {author} {\bibinfo {author} {\bibfnamefont {F.}~\bibnamefont
  {Ercolessi}}\ and\ \bibinfo {author} {\bibfnamefont {J.}~\bibnamefont
  {Adams}},\ }\href {\doibase 10.1209/0295-5075/26/8/005} {\bibfield  {journal}
  {\bibinfo  {journal} {Europhys. Lett.}\ }\textbf {\bibinfo {volume} {26}},\
  \bibinfo {pages} {583} (\bibinfo {year} {1994})}\BibitemShut {NoStop}%
\bibitem [{\citenamefont {Jaramillo-Botero}\ \emph {et~al.}(2014)\citenamefont
  {Jaramillo-Botero}, \citenamefont {Naserifar},\ and\ \citenamefont
  {Goddard}}]{jaramillo-botero_general_2014}%
  \BibitemOpen
  \bibfield  {author} {\bibinfo {author} {\bibfnamefont {A.}~\bibnamefont
  {Jaramillo-Botero}}, \bibinfo {author} {\bibfnamefont {S.}~\bibnamefont
  {Naserifar}}, \ and\ \bibinfo {author} {\bibfnamefont {W.}~\bibnamefont
  {Goddard}},\ }\href {\doibase 10.1021/ct5001044} {\bibfield  {journal}
  {\bibinfo  {journal} {J. Chem. Theory Comput.}\ }\textbf {\bibinfo {volume}
  {10}},\ \bibinfo {pages} {1426} (\bibinfo {year} {2014})}\BibitemShut
  {NoStop}%
\bibitem [{\citenamefont {Cs\'{a}nyi}\ \emph {et~al.}(2004)\citenamefont
  {Cs\'{a}nyi}, \citenamefont {Albaret}, \citenamefont {Payne},\ and\
  \citenamefont {De~Vita}}]{csanyi_learn_2004}%
  \BibitemOpen
  \bibfield  {author} {\bibinfo {author} {\bibfnamefont {G.}~\bibnamefont
  {Cs\'{a}nyi}}, \bibinfo {author} {\bibfnamefont {T.}~\bibnamefont {Albaret}},
  \bibinfo {author} {\bibfnamefont {M.}~\bibnamefont {Payne}}, \ and\ \bibinfo
  {author} {\bibfnamefont {A.}~\bibnamefont {De~Vita}},\ }\href {\doibase
  10.1103/PhysRevLett.93.175503} {\bibfield  {journal} {\bibinfo  {journal}
  {Phys. Rev. Lett.}\ }\textbf {\bibinfo {volume} {93}},\ \bibinfo {pages}
  {175503} (\bibinfo {year} {2004})}\BibitemShut {NoStop}%
\bibitem [{\citenamefont {Moras}\ \emph {et~al.}(2010)\citenamefont {Moras},
  \citenamefont {Ciacchi}, \citenamefont {Els\"{a}sser}, \citenamefont
  {Gumbsch},\ and\ \citenamefont {De~Vita}}]{moras_atomically_2010}%
  \BibitemOpen
  \bibfield  {author} {\bibinfo {author} {\bibfnamefont {G.}~\bibnamefont
  {Moras}}, \bibinfo {author} {\bibfnamefont {L.}~\bibnamefont {Ciacchi}},
  \bibinfo {author} {\bibfnamefont {C.}~\bibnamefont {Els\"{a}sser}}, \bibinfo
  {author} {\bibfnamefont {P.}~\bibnamefont {Gumbsch}}, \ and\ \bibinfo
  {author} {\bibfnamefont {A.}~\bibnamefont {De~Vita}},\ }\href {\doibase
  10.1103/PhysRevLett.105.075502} {\bibfield  {journal} {\bibinfo  {journal}
  {Phys. Rev. Lett.}\ }\textbf {\bibinfo {volume} {105}},\ \bibinfo {pages}
  {075502} (\bibinfo {year} {2010})}\BibitemShut {NoStop}%
\bibitem [{\citenamefont {Kermode}\ \emph {et~al.}(2013)\citenamefont
  {Kermode}, \citenamefont {Ben-Bashat}, \citenamefont {Atrash}, \citenamefont
  {Cilliers}, \citenamefont {Sherman},\ and\ \citenamefont
  {De~Vita}}]{kermode_macroscopic_2013}%
  \BibitemOpen
  \bibfield  {author} {\bibinfo {author} {\bibfnamefont {J.}~\bibnamefont
  {Kermode}}, \bibinfo {author} {\bibfnamefont {L.}~\bibnamefont {Ben-Bashat}},
  \bibinfo {author} {\bibfnamefont {F.}~\bibnamefont {Atrash}}, \bibinfo
  {author} {\bibfnamefont {J.}~\bibnamefont {Cilliers}}, \bibinfo {author}
  {\bibfnamefont {D.}~\bibnamefont {Sherman}}, \ and\ \bibinfo {author}
  {\bibfnamefont {A.}~\bibnamefont {De~Vita}},\ }\href {\doibase
  10.1038/ncomms3441} {\bibfield  {journal} {\bibinfo  {journal} {Nat.
  Commun.}\ }\textbf {\bibinfo {volume} {4}},\ \bibinfo {pages} {2441}
  (\bibinfo {year} {2013})}\BibitemShut {NoStop}%
\bibitem [{\citenamefont {Avinun}\ \emph {et~al.}(1998)\citenamefont {Avinun},
  \citenamefont {Barel}, \citenamefont {Kaplan}, \citenamefont {Eizenberg},
  \citenamefont {Naik}, \citenamefont {Guo}, \citenamefont {Chen},
  \citenamefont {Mosely}, \citenamefont {Littau}, \citenamefont {Zhou},\ and\
  \citenamefont {Chen}}]{avinun_nucleation_1998}%
  \BibitemOpen
  \bibfield  {author} {\bibinfo {author} {\bibfnamefont {M.}~\bibnamefont
  {Avinun}}, \bibinfo {author} {\bibfnamefont {N.}~\bibnamefont {Barel}},
  \bibinfo {author} {\bibfnamefont {W.}~\bibnamefont {Kaplan}}, \bibinfo
  {author} {\bibfnamefont {M.}~\bibnamefont {Eizenberg}}, \bibinfo {author}
  {\bibfnamefont {M.}~\bibnamefont {Naik}}, \bibinfo {author} {\bibfnamefont
  {T.}~\bibnamefont {Guo}}, \bibinfo {author} {\bibfnamefont {L.}~\bibnamefont
  {Chen}}, \bibinfo {author} {\bibfnamefont {R.}~\bibnamefont {Mosely}},
  \bibinfo {author} {\bibfnamefont {K.}~\bibnamefont {Littau}}, \bibinfo
  {author} {\bibfnamefont {S.}~\bibnamefont {Zhou}}, \ and\ \bibinfo {author}
  {\bibfnamefont {L.}~\bibnamefont {Chen}},\ }\href {\doibase
  10.1016/S0040-6090(97)01067-5} {\bibfield  {journal} {\bibinfo  {journal}
  {Thin Solid Films}\ }\textbf {\bibinfo {volume} {320}},\ \bibinfo {pages}
  {67} (\bibinfo {year} {1998})}\BibitemShut {NoStop}%
\bibitem [{\citenamefont {Chun}\ \emph {et~al.}(2001)\citenamefont {Chun},
  \citenamefont {Desjardins}, \citenamefont {Lavoie}, \citenamefont {Shin},
  \citenamefont {Cabral}, \citenamefont {Petrov},\ and\ \citenamefont
  {Greene}}]{chun_interfacial_2001-1}%
  \BibitemOpen
  \bibfield  {author} {\bibinfo {author} {\bibfnamefont {J.-S.}\ \bibnamefont
  {Chun}}, \bibinfo {author} {\bibfnamefont {P.}~\bibnamefont {Desjardins}},
  \bibinfo {author} {\bibfnamefont {C.}~\bibnamefont {Lavoie}}, \bibinfo
  {author} {\bibfnamefont {C.-S.}\ \bibnamefont {Shin}}, \bibinfo {author}
  {\bibfnamefont {C.}~\bibnamefont {Cabral}}, \bibinfo {author} {\bibfnamefont
  {I.}~\bibnamefont {Petrov}}, \ and\ \bibinfo {author} {\bibfnamefont
  {J.}~\bibnamefont {Greene}},\ }\href {\doibase 10.1063/1.1372162} {\bibfield
  {journal} {\bibinfo  {journal} {J. Appl. Phys.}\ }\textbf {\bibinfo {volume}
  {89}},\ \bibinfo {pages} {7841} (\bibinfo {year} {2001})}\BibitemShut
  {NoStop}%
\bibitem [{\citenamefont {Howe}(1993{\natexlab{b}})}]{howe_bonding_1993-1}%
  \BibitemOpen
  \bibfield  {author} {\bibinfo {author} {\bibfnamefont {J.}~\bibnamefont
  {Howe}},\ }\href@noop {} {\bibfield  {journal} {\bibinfo  {journal} {Int.
  Mater. Rev.}\ }\textbf {\bibinfo {volume} {38}},\ \bibinfo {pages} {257}
  (\bibinfo {year} {1993}{\natexlab{b}})}\BibitemShut {NoStop}%
\bibitem [{\citenamefont {Ernst}(1995)}]{ernst_metal-oxide_1995}%
  \BibitemOpen
  \bibfield  {author} {\bibinfo {author} {\bibfnamefont {F.}~\bibnamefont
  {Ernst}},\ }\href {\doibase 10.1016/0927-796X(95)80001-8} {\bibfield
  {journal} {\bibinfo  {journal} {Mater. Sci. Eng., R}\ }\textbf {\bibinfo
  {volume} {14}},\ \bibinfo {pages} {97} (\bibinfo {year} {1995})}\BibitemShut
  {NoStop}%
\bibitem [{\citenamefont {Liu}\ \emph {et~al.}(2004)\citenamefont {Liu},
  \citenamefont {Wang},\ and\ \citenamefont {Ye}}]{liu_first-principles_2004}%
  \BibitemOpen
  \bibfield  {author} {\bibinfo {author} {\bibfnamefont {L.}~\bibnamefont
  {Liu}}, \bibinfo {author} {\bibfnamefont {S.}~\bibnamefont {Wang}}, \ and\
  \bibinfo {author} {\bibfnamefont {H.}~\bibnamefont {Ye}},\ }\href {\doibase
  10.1016/j.actamat.2004.04.022} {\bibfield  {journal} {\bibinfo  {journal}
  {Acta Mater.}\ }\textbf {\bibinfo {volume} {52}},\ \bibinfo {pages} {3681}
  (\bibinfo {year} {2004})}\BibitemShut {NoStop}%
\bibitem [{\citenamefont {Song}\ and\ \citenamefont
  {Srolovitz}(2006)}]{song_adhesion_2006}%
  \BibitemOpen
  \bibfield  {author} {\bibinfo {author} {\bibfnamefont {J.}~\bibnamefont
  {Song}}\ and\ \bibinfo {author} {\bibfnamefont {D.}~\bibnamefont
  {Srolovitz}},\ }\href {\doibase 10.1016/j.actamat.2006.07.011} {\bibfield
  {journal} {\bibinfo  {journal} {Acta Mater.}\ }\textbf {\bibinfo {volume}
  {54}},\ \bibinfo {pages} {5305} (\bibinfo {year} {2006})}\BibitemShut
  {NoStop}%
\bibitem [{\citenamefont {Zhang}\ \emph {et~al.}(2007)\citenamefont {Zhang},
  \citenamefont {Liu},\ and\ \citenamefont
  {Wang}}]{zhang_first-principles_2007}%
  \BibitemOpen
  \bibfield  {author} {\bibinfo {author} {\bibfnamefont {H.}~\bibnamefont
  {Zhang}}, \bibinfo {author} {\bibfnamefont {L.}~\bibnamefont {Liu}}, \ and\
  \bibinfo {author} {\bibfnamefont {S.}~\bibnamefont {Wang}},\ }\href {\doibase
  10.1016/j.commatsci.2006.05.017} {\bibfield  {journal} {\bibinfo  {journal}
  {Comput. Mater. Sci.}\ }\textbf {\bibinfo {volume} {38}},\ \bibinfo {pages}
  {800} (\bibinfo {year} {2007})}\BibitemShut {NoStop}%
\bibitem [{\citenamefont {Zhang}\ and\ \citenamefont
  {Wang}(2007)}]{zhang_effects_2007}%
  \BibitemOpen
  \bibfield  {author} {\bibinfo {author} {\bibfnamefont {H.}~\bibnamefont
  {Zhang}}\ and\ \bibinfo {author} {\bibfnamefont {S.}~\bibnamefont {Wang}},\
  }\href {\doibase 10.1088/0953-8984/19/22/226003} {\bibfield  {journal}
  {\bibinfo  {journal} {J. Phys.: Condens. Matter}\ }\textbf {\bibinfo {volume}
  {19}},\ \bibinfo {pages} {226003} (\bibinfo {year} {2007})}\BibitemShut
  {NoStop}%
\bibitem [{\citenamefont {Yadav}\ \emph {et~al.}(2012)\citenamefont {Yadav},
  \citenamefont {Ramprasad}, \citenamefont {Misra},\ and\ \citenamefont
  {Liu}}]{yadav_first-principles_2012}%
  \BibitemOpen
  \bibfield  {author} {\bibinfo {author} {\bibfnamefont {S.}~\bibnamefont
  {Yadav}}, \bibinfo {author} {\bibfnamefont {R.}~\bibnamefont {Ramprasad}},
  \bibinfo {author} {\bibfnamefont {A.}~\bibnamefont {Misra}}, \ and\ \bibinfo
  {author} {\bibfnamefont {X.-Y.}\ \bibnamefont {Liu}},\ }\href {\doibase
  10.1063/1.3703663} {\bibfield  {journal} {\bibinfo  {journal} {J. Appl.
  Phys.}\ }\textbf {\bibinfo {volume} {111}},\ \bibinfo {pages} {083505}
  (\bibinfo {year} {2012})}\BibitemShut {NoStop}%
\bibitem [{\citenamefont {Yadav}\ \emph {et~al.}(2014)\citenamefont {Yadav},
  \citenamefont {Ramprasad}, \citenamefont {Wang}, \citenamefont {Misra},\ and\
  \citenamefont {Liu}}]{yadav_first-principles_2014}%
  \BibitemOpen
  \bibfield  {author} {\bibinfo {author} {\bibfnamefont {S.}~\bibnamefont
  {Yadav}}, \bibinfo {author} {\bibfnamefont {R.}~\bibnamefont {Ramprasad}},
  \bibinfo {author} {\bibfnamefont {J.}~\bibnamefont {Wang}}, \bibinfo {author}
  {\bibfnamefont {A.}~\bibnamefont {Misra}}, \ and\ \bibinfo {author}
  {\bibfnamefont {X.-Y.}\ \bibnamefont {Liu}},\ }\href {\doibase
  10.1088/0965-0393/22/3/035020} {\bibfield  {journal} {\bibinfo  {journal}
  {Modell. Simul. Mater. Sci. Eng.}\ }\textbf {\bibinfo {volume} {22}},\
  \bibinfo {pages} {035020} (\bibinfo {year} {2014})}\BibitemShut {NoStop}%
\bibitem [{\citenamefont {Yadav}\ \emph {et~al.}(2015)\citenamefont {Yadav},
  \citenamefont {Shao}, \citenamefont {Wang},\ and\ \citenamefont
  {Liu}}]{yadav_structural_2015}%
  \BibitemOpen
  \bibfield  {author} {\bibinfo {author} {\bibfnamefont {S.}~\bibnamefont
  {Yadav}}, \bibinfo {author} {\bibfnamefont {S.}~\bibnamefont {Shao}},
  \bibinfo {author} {\bibfnamefont {J.}~\bibnamefont {Wang}}, \ and\ \bibinfo
  {author} {\bibfnamefont {X.-Y.}\ \bibnamefont {Liu}},\ }\href {\doibase
  10.1038/srep17380} {\bibfield  {journal} {\bibinfo  {journal} {Sci. Rep.}\
  }\textbf {\bibinfo {volume} {5}},\ \bibinfo {pages} {17380} (\bibinfo {year}
  {2015})}\BibitemShut {NoStop}%
\bibitem [{\citenamefont {Li}\ \emph {et~al.}(2015)\citenamefont {Li},
  \citenamefont {Yadav}, \citenamefont {Wang}, \citenamefont {Liu},\ and\
  \citenamefont {Misra}}]{li_growth_2015-1}%
  \BibitemOpen
  \bibfield  {author} {\bibinfo {author} {\bibfnamefont {N.}~\bibnamefont
  {Li}}, \bibinfo {author} {\bibfnamefont {S.}~\bibnamefont {Yadav}}, \bibinfo
  {author} {\bibfnamefont {J.}~\bibnamefont {Wang}}, \bibinfo {author}
  {\bibfnamefont {X.-Y.}\ \bibnamefont {Liu}}, \ and\ \bibinfo {author}
  {\bibfnamefont {A.}~\bibnamefont {Misra}},\ }\href {\doibase
  10.1038/srep18554} {\bibfield  {journal} {\bibinfo  {journal} {Sci. Rep.}\
  }\textbf {\bibinfo {volume} {5}},\ \bibinfo {pages} {18554} (\bibinfo {year}
  {2015})}\BibitemShut {NoStop}%
\bibitem [{\citenamefont {Lin}\ \emph {et~al.}(2017)\citenamefont {Lin},
  \citenamefont {Peng}, \citenamefont {Fu}, \citenamefont {Zhao}, \citenamefont
  {Feng}, \citenamefont {Huang},\ and\ \citenamefont {Wang}}]{lin_atomic_2017}%
  \BibitemOpen
  \bibfield  {author} {\bibinfo {author} {\bibfnamefont {Z.}~\bibnamefont
  {Lin}}, \bibinfo {author} {\bibfnamefont {X.}~\bibnamefont {Peng}}, \bibinfo
  {author} {\bibfnamefont {T.}~\bibnamefont {Fu}}, \bibinfo {author}
  {\bibfnamefont {Y.}~\bibnamefont {Zhao}}, \bibinfo {author} {\bibfnamefont
  {C.}~\bibnamefont {Feng}}, \bibinfo {author} {\bibfnamefont {C.}~\bibnamefont
  {Huang}}, \ and\ \bibinfo {author} {\bibfnamefont {Z.}~\bibnamefont {Wang}},\
  }\href {\doibase 10.1016/j.physe.2017.01.025} {\bibfield  {journal} {\bibinfo
   {journal} {Phys. E}\ }\textbf {\bibinfo {volume} {89}},\ \bibinfo {pages}
  {15} (\bibinfo {year} {2017})}\BibitemShut {NoStop}%
\bibitem [{\citenamefont {Liu}\ \emph {et~al.}(2005)\citenamefont {Liu},
  \citenamefont {Wang},\ and\ \citenamefont {Ye}}]{liu_first-principles_2005}%
  \BibitemOpen
  \bibfield  {author} {\bibinfo {author} {\bibfnamefont {L.}~\bibnamefont
  {Liu}}, \bibinfo {author} {\bibfnamefont {S.}~\bibnamefont {Wang}}, \ and\
  \bibinfo {author} {\bibfnamefont {H.}~\bibnamefont {Ye}},\ }\href {\doibase
  10.1088/0953-8984/17/35/002} {\bibfield  {journal} {\bibinfo  {journal} {J.
  Phys.: Condens. Matter}\ }\textbf {\bibinfo {volume} {17}},\ \bibinfo {pages}
  {5335} (\bibinfo {year} {2005})}\BibitemShut {NoStop}%
\bibitem [{\citenamefont {Jacobsen}\ \emph {et~al.}(1995)\citenamefont
  {Jacobsen}, \citenamefont {Hammer}, \citenamefont {Jacobsen},\ and\
  \citenamefont {N{\o}rskov}}]{jacobsen_electronic_1995}%
  \BibitemOpen
  \bibfield  {author} {\bibinfo {author} {\bibfnamefont {J.}~\bibnamefont
  {Jacobsen}}, \bibinfo {author} {\bibfnamefont {B.}~\bibnamefont {Hammer}},
  \bibinfo {author} {\bibfnamefont {K.~W.}\ \bibnamefont {Jacobsen}}, \ and\
  \bibinfo {author} {\bibfnamefont {J.}~\bibnamefont {N{\o}rskov}},\ }\href
  {\doibase 10.1103/PhysRevB.52.14954} {\bibfield  {journal} {\bibinfo
  {journal} {Phys. Rev. B}\ }\textbf {\bibinfo {volume} {52}},\ \bibinfo
  {pages} {14954} (\bibinfo {year} {1995})}\BibitemShut {NoStop}%
\bibitem [{\citenamefont {Jennison}\ and\ \citenamefont
  {Bogicevic}(2000)}]{jennison_ultrathin_2000}%
  \BibitemOpen
  \bibfield  {author} {\bibinfo {author} {\bibfnamefont {D.}~\bibnamefont
  {Jennison}}\ and\ \bibinfo {author} {\bibfnamefont {A.}~\bibnamefont
  {Bogicevic}},\ }\href {\doibase 10.1016/S0039-6028(00)00578-1} {\bibfield
  {journal} {\bibinfo  {journal} {Surf. Sci.}\ }\textbf {\bibinfo {volume}
  {464}},\ \bibinfo {pages} {108} (\bibinfo {year} {2000})}\BibitemShut
  {NoStop}%
\bibitem [{\citenamefont {Schaich}\ \emph {et~al.}(1997)\citenamefont
  {Schaich}, \citenamefont {Braun}, \citenamefont {Toennies}, \citenamefont
  {Buck},\ and\ \citenamefont {W\"{o}ll}}]{schaich_structural_1997}%
  \BibitemOpen
  \bibfield  {author} {\bibinfo {author} {\bibfnamefont {T.}~\bibnamefont
  {Schaich}}, \bibinfo {author} {\bibfnamefont {J.}~\bibnamefont {Braun}},
  \bibinfo {author} {\bibfnamefont {J.}~\bibnamefont {Toennies}}, \bibinfo
  {author} {\bibfnamefont {M.}~\bibnamefont {Buck}}, \ and\ \bibinfo {author}
  {\bibfnamefont {C.}~\bibnamefont {W\"{o}ll}},\ }\href {\doibase
  10.1016/S0039-6028(97)00348-8} {\bibfield  {journal} {\bibinfo  {journal}
  {Surf. Sci.}\ }\textbf {\bibinfo {volume} {385}},\ \bibinfo {pages} {L958}
  (\bibinfo {year} {1997})}\BibitemShut {NoStop}%
\bibitem [{\citenamefont {Scholze}\ \emph {et~al.}(1996)\citenamefont
  {Scholze}, \citenamefont {Schmidt},\ and\ \citenamefont
  {Bechstedt}}]{scholze_structure_1996}%
  \BibitemOpen
  \bibfield  {author} {\bibinfo {author} {\bibfnamefont {A.}~\bibnamefont
  {Scholze}}, \bibinfo {author} {\bibfnamefont {W.}~\bibnamefont {Schmidt}}, \
  and\ \bibinfo {author} {\bibfnamefont {F.}~\bibnamefont {Bechstedt}},\ }\href
  {\doibase 10.1103/PhysRevB.53.13725} {\bibfield  {journal} {\bibinfo
  {journal} {Phys. Rev. B}\ }\textbf {\bibinfo {volume} {53}},\ \bibinfo
  {pages} {13725} (\bibinfo {year} {1996})}\BibitemShut {NoStop}%
\bibitem [{\citenamefont {Kern}\ \emph
  {et~al.}(1996{\natexlab{a}})\citenamefont {Kern}, \citenamefont {Hafner},\
  and\ \citenamefont {Kresse}}]{kern_atomic_1996}%
  \BibitemOpen
  \bibfield  {author} {\bibinfo {author} {\bibfnamefont {G.}~\bibnamefont
  {Kern}}, \bibinfo {author} {\bibfnamefont {J.}~\bibnamefont {Hafner}}, \ and\
  \bibinfo {author} {\bibfnamefont {G.}~\bibnamefont {Kresse}},\ }\href
  {\doibase 10.1016/0039-6028(96)00837-0} {\bibfield  {journal} {\bibinfo
  {journal} {Surf. Sci.}\ }\textbf {\bibinfo {volume} {366}},\ \bibinfo {pages}
  {445} (\bibinfo {year} {1996}{\natexlab{a}})}\BibitemShut {NoStop}%
\bibitem [{\citenamefont {Kern}\ \emph
  {et~al.}(1996{\natexlab{b}})\citenamefont {Kern}, \citenamefont {Hafner},\
  and\ \citenamefont {Kresse}}]{kern_atomic_1996-1}%
  \BibitemOpen
  \bibfield  {author} {\bibinfo {author} {\bibfnamefont {G.}~\bibnamefont
  {Kern}}, \bibinfo {author} {\bibfnamefont {J.}~\bibnamefont {Hafner}}, \ and\
  \bibinfo {author} {\bibfnamefont {G.}~\bibnamefont {Kresse}},\ }\href
  {\doibase 10.1016/0039-6028(96)00836-9} {\bibfield  {journal} {\bibinfo
  {journal} {Surf. Sci.}\ }\textbf {\bibinfo {volume} {366}},\ \bibinfo {pages}
  {464} (\bibinfo {year} {1996}{\natexlab{b}})}\BibitemShut {NoStop}%
\bibitem [{\citenamefont {Kern}\ \emph {et~al.}(1997)\citenamefont {Kern},
  \citenamefont {Hafner},\ and\ \citenamefont {Kresse}}]{kern_atomic_1997}%
  \BibitemOpen
  \bibfield  {author} {\bibinfo {author} {\bibfnamefont {G.}~\bibnamefont
  {Kern}}, \bibinfo {author} {\bibfnamefont {J.}~\bibnamefont {Hafner}}, \ and\
  \bibinfo {author} {\bibfnamefont {G.}~\bibnamefont {Kresse}},\ }\href
  {\doibase 10.1016/S0039-6028(97)00168-4} {\bibfield  {journal} {\bibinfo
  {journal} {Surf. Sci.}\ }\textbf {\bibinfo {volume} {384}},\ \bibinfo {pages}
  {94} (\bibinfo {year} {1997})}\BibitemShut {NoStop}%
\bibitem [{\citenamefont {Pepper}(1982)}]{pepper_effect_1982}%
  \BibitemOpen
  \bibfield  {author} {\bibinfo {author} {\bibfnamefont {S.}~\bibnamefont
  {Pepper}},\ }\href {\doibase 10.1116/1.571616} {\bibfield  {journal}
  {\bibinfo  {journal} {J. Vac. Sci. Technol.}\ }\textbf {\bibinfo {volume}
  {20}},\ \bibinfo {pages} {643} (\bibinfo {year} {1982})}\BibitemShut
  {NoStop}%
\bibitem [{\citenamefont {Zhu}\ and\ \citenamefont
  {Fang}(2016)}]{zhu_study_2016}%
  \BibitemOpen
  \bibfield  {author} {\bibinfo {author} {\bibfnamefont {P.}~\bibnamefont
  {Zhu}}\ and\ \bibinfo {author} {\bibfnamefont {F.}~\bibnamefont {Fang}},\
  }\href {\doibase 10.1016/j.commatsci.2016.03.023} {\bibfield  {journal}
  {\bibinfo  {journal} {Comput. Mater. Sci.}\ }\textbf {\bibinfo {volume}
  {118}},\ \bibinfo {pages} {192} (\bibinfo {year} {2016})}\BibitemShut
  {NoStop}%
\bibitem [{\citenamefont {Kresse}\ and\ \citenamefont
  {Hafner}(1993)}]{kresse_ab_1993}%
  \BibitemOpen
  \bibfield  {author} {\bibinfo {author} {\bibfnamefont {G.}~\bibnamefont
  {Kresse}}\ and\ \bibinfo {author} {\bibfnamefont {J.}~\bibnamefont
  {Hafner}},\ }\href {\doibase 10.1103/PhysRevB.47.558} {\bibfield  {journal}
  {\bibinfo  {journal} {Phys. Rev. B}\ }\textbf {\bibinfo {volume} {47}},\
  \bibinfo {pages} {558} (\bibinfo {year} {1993})}\BibitemShut {NoStop}%
\bibitem [{\citenamefont {Kresse}\ and\ \citenamefont
  {Hafner}(1994)}]{kresse_ab_1994}%
  \BibitemOpen
  \bibfield  {author} {\bibinfo {author} {\bibfnamefont {G.}~\bibnamefont
  {Kresse}}\ and\ \bibinfo {author} {\bibfnamefont {J.}~\bibnamefont
  {Hafner}},\ }\href {\doibase 10.1103/PhysRevB.49.14251} {\bibfield  {journal}
  {\bibinfo  {journal} {Phys. Rev. B}\ }\textbf {\bibinfo {volume} {49}},\
  \bibinfo {pages} {14251} (\bibinfo {year} {1994})}\BibitemShut {NoStop}%
\bibitem [{\citenamefont {Kresse}\ and\ \citenamefont
  {Furthm\"{u}ller}(1996{\natexlab{a}})}]{kresse_efficient_1996}%
  \BibitemOpen
  \bibfield  {author} {\bibinfo {author} {\bibfnamefont {G.}~\bibnamefont
  {Kresse}}\ and\ \bibinfo {author} {\bibfnamefont {J.}~\bibnamefont
  {Furthm\"{u}ller}},\ }\href {\doibase 10.1103/PhysRevB.54.11169} {\bibfield
  {journal} {\bibinfo  {journal} {Phys. Rev. B}\ }\textbf {\bibinfo {volume}
  {54}},\ \bibinfo {pages} {11169} (\bibinfo {year}
  {1996}{\natexlab{a}})}\BibitemShut {NoStop}%
\bibitem [{\citenamefont {Kresse}\ and\ \citenamefont
  {Furthm\"{u}ller}(1996{\natexlab{b}})}]{kresse_efficiency_1996}%
  \BibitemOpen
  \bibfield  {author} {\bibinfo {author} {\bibfnamefont {G.}~\bibnamefont
  {Kresse}}\ and\ \bibinfo {author} {\bibfnamefont {J.}~\bibnamefont
  {Furthm\"{u}ller}},\ }\href {\doibase 10.1016/0927-0256(96)00008-0}
  {\bibfield  {journal} {\bibinfo  {journal} {Comput. Mater. Sci.}\ }\textbf
  {\bibinfo {volume} {6}},\ \bibinfo {pages} {15} (\bibinfo {year}
  {1996}{\natexlab{b}})}\BibitemShut {NoStop}%
\bibitem [{\citenamefont {Bl\"{o}chl}(1994)}]{blochl_projector_1994}%
  \BibitemOpen
  \bibfield  {author} {\bibinfo {author} {\bibfnamefont {P.}~\bibnamefont
  {Bl\"{o}chl}},\ }\href {\doibase 10.1103/PhysRevB.50.17953} {\bibfield
  {journal} {\bibinfo  {journal} {Phys. Rev. B}\ }\textbf {\bibinfo {volume}
  {50}},\ \bibinfo {pages} {17953} (\bibinfo {year} {1994})}\BibitemShut
  {NoStop}%
\bibitem [{\citenamefont {Kresse}\ and\ \citenamefont
  {Joubert}(1999)}]{kresse_ultrasoft_1999}%
  \BibitemOpen
  \bibfield  {author} {\bibinfo {author} {\bibfnamefont {G.}~\bibnamefont
  {Kresse}}\ and\ \bibinfo {author} {\bibfnamefont {D.}~\bibnamefont
  {Joubert}},\ }\href {\doibase 10.1103/PhysRevB.59.1758} {\bibfield  {journal}
  {\bibinfo  {journal} {Phys. Rev. B}\ }\textbf {\bibinfo {volume} {59}},\
  \bibinfo {pages} {1758} (\bibinfo {year} {1999})}\BibitemShut {NoStop}%
\bibitem [{\citenamefont {Perdew}\ \emph {et~al.}(1996)\citenamefont {Perdew},
  \citenamefont {Burke},\ and\ \citenamefont
  {Ernzerhof}}]{perdew_generalized_1996}%
  \BibitemOpen
  \bibfield  {author} {\bibinfo {author} {\bibfnamefont {J.}~\bibnamefont
  {Perdew}}, \bibinfo {author} {\bibfnamefont {K.}~\bibnamefont {Burke}}, \
  and\ \bibinfo {author} {\bibfnamefont {M.}~\bibnamefont {Ernzerhof}},\ }\href
  {\doibase 10.1103/PhysRevLett.77.3865} {\bibfield  {journal} {\bibinfo
  {journal} {Phys. Rev. Lett.}\ }\textbf {\bibinfo {volume} {77}},\ \bibinfo
  {pages} {3865} (\bibinfo {year} {1996})}\BibitemShut {NoStop}%
\bibitem [{\citenamefont {Stampfl}\ and\ \citenamefont {Van~de
  Walle}(1999)}]{stampfl_density-functional_1999}%
  \BibitemOpen
  \bibfield  {author} {\bibinfo {author} {\bibfnamefont {C.}~\bibnamefont
  {Stampfl}}\ and\ \bibinfo {author} {\bibfnamefont {C.}~\bibnamefont {Van~de
  Walle}},\ }\href {\doibase 10.1103/PhysRevB.59.5521} {\bibfield  {journal}
  {\bibinfo  {journal} {Phys. Rev. B}\ }\textbf {\bibinfo {volume} {59}},\
  \bibinfo {pages} {5521} (\bibinfo {year} {1999})}\BibitemShut {NoStop}%
\bibitem [{\citenamefont {Perdew}\ and\ \citenamefont
  {Zunger}(1981)}]{perdew_self-interaction_1981}%
  \BibitemOpen
  \bibfield  {author} {\bibinfo {author} {\bibfnamefont {J.}~\bibnamefont
  {Perdew}}\ and\ \bibinfo {author} {\bibfnamefont {A.}~\bibnamefont
  {Zunger}},\ }\href {\doibase 10.1103/PhysRevB.23.5048} {\bibfield  {journal}
  {\bibinfo  {journal} {Phys. Rev. B}\ }\textbf {\bibinfo {volume} {23}},\
  \bibinfo {pages} {5048} (\bibinfo {year} {1981})}\BibitemShut {NoStop}%
\bibitem [{\citenamefont {van~de Walle}\ and\ \citenamefont
  {Ceder}(1999)}]{van_de_walle_correcting_1999}%
  \BibitemOpen
  \bibfield  {author} {\bibinfo {author} {\bibfnamefont {A.}~\bibnamefont
  {van~de Walle}}\ and\ \bibinfo {author} {\bibfnamefont {G.}~\bibnamefont
  {Ceder}},\ }\href {\doibase 10.1103/PhysRevB.59.14992} {\bibfield  {journal}
  {\bibinfo  {journal} {Phys. Rev. B}\ }\textbf {\bibinfo {volume} {59}},\
  \bibinfo {pages} {14992} (\bibinfo {year} {1999})}\BibitemShut {NoStop}%
\bibitem [{\citenamefont {Grabowski}\ \emph {et~al.}(2007)\citenamefont
  {Grabowski}, \citenamefont {Hickel},\ and\ \citenamefont
  {Neugebauer}}]{grabowski_ab_2007}%
  \BibitemOpen
  \bibfield  {author} {\bibinfo {author} {\bibfnamefont {B.}~\bibnamefont
  {Grabowski}}, \bibinfo {author} {\bibfnamefont {T.}~\bibnamefont {Hickel}}, \
  and\ \bibinfo {author} {\bibfnamefont {J.}~\bibnamefont {Neugebauer}},\
  }\href {\doibase 10.1103/PhysRevB.76.024309} {\bibfield  {journal} {\bibinfo
  {journal} {Phys. Rev. B}\ }\textbf {\bibinfo {volume} {76}},\ \bibinfo
  {pages} {024309} (\bibinfo {year} {2007})}\BibitemShut {NoStop}%
\bibitem [{\citenamefont {Csonka}\ \emph {et~al.}(2009)\citenamefont {Csonka},
  \citenamefont {Perdew}, \citenamefont {Ruzsinszky}, \citenamefont
  {Philipsen}, \citenamefont {Leb\`egue}, \citenamefont {Paier}, \citenamefont
  {Vydrov},\ and\ \citenamefont {\'Angy\'an}}]{csonka_assessing_2009}%
  \BibitemOpen
  \bibfield  {author} {\bibinfo {author} {\bibfnamefont {G.~I.}\ \bibnamefont
  {Csonka}}, \bibinfo {author} {\bibfnamefont {J.~P.}\ \bibnamefont {Perdew}},
  \bibinfo {author} {\bibfnamefont {A.}~\bibnamefont {Ruzsinszky}}, \bibinfo
  {author} {\bibfnamefont {P.~H.~T.}\ \bibnamefont {Philipsen}}, \bibinfo
  {author} {\bibfnamefont {S.}~\bibnamefont {Leb\`egue}}, \bibinfo {author}
  {\bibfnamefont {J.}~\bibnamefont {Paier}}, \bibinfo {author} {\bibfnamefont
  {O.~A.}\ \bibnamefont {Vydrov}}, \ and\ \bibinfo {author} {\bibfnamefont
  {J.~G.}\ \bibnamefont {\'Angy\'an}},\ }\href {\doibase
  10.1103/PhysRevB.79.155107} {\bibfield  {journal} {\bibinfo  {journal} {Phys.
  Rev. B}\ }\textbf {\bibinfo {volume} {79}},\ \bibinfo {pages} {155107}
  (\bibinfo {year} {2009})}\BibitemShut {NoStop}%
\bibitem [{\citenamefont {Michaelides}\ and\ \citenamefont
  {Scheffler}(2014)}]{wandelt_introduction_2014}%
  \BibitemOpen
  \bibfield  {author} {\bibinfo {author} {\bibfnamefont {A.}~\bibnamefont
  {Michaelides}}\ and\ \bibinfo {author} {\bibfnamefont {M.}~\bibnamefont
  {Scheffler}},\ }in\ \href@noop {} {\emph {\bibinfo {booktitle} {Surface and
  {Interface} {Science}}}},\ \bibinfo {editor} {edited by\ \bibinfo {editor}
  {\bibfnamefont {K.}~\bibnamefont {Wandelt}}}\ (\bibinfo  {publisher}
  {Wiley-VCH Verlag GmbH \& Co. KGaA},\ \bibinfo {address} {Weinheim,
  Germany},\ \bibinfo {year} {2014})\ pp.\ \bibinfo {pages}
  {13--72}\BibitemShut {NoStop}%
\bibitem [{\citenamefont {Monkhorst}\ and\ \citenamefont
  {Pack}(1976)}]{monkhorst_special_1976}%
  \BibitemOpen
  \bibfield  {author} {\bibinfo {author} {\bibfnamefont {H.}~\bibnamefont
  {Monkhorst}}\ and\ \bibinfo {author} {\bibfnamefont {J.}~\bibnamefont
  {Pack}},\ }\href {\doibase 10.1103/PhysRevB.13.5188} {\bibfield  {journal}
  {\bibinfo  {journal} {Phys. Rev. B}\ }\textbf {\bibinfo {volume} {13}},\
  \bibinfo {pages} {5188} (\bibinfo {year} {1976})}\BibitemShut {NoStop}%
\bibitem [{\citenamefont {Methfessel}\ and\ \citenamefont
  {Paxton}(1989)}]{methfessel_high-precision_1989}%
  \BibitemOpen
  \bibfield  {author} {\bibinfo {author} {\bibfnamefont {M.}~\bibnamefont
  {Methfessel}}\ and\ \bibinfo {author} {\bibfnamefont {A.}~\bibnamefont
  {Paxton}},\ }\href {\doibase 10.1103/PhysRevB.40.3616} {\bibfield  {journal}
  {\bibinfo  {journal} {Phys. Rev. B}\ }\textbf {\bibinfo {volume} {40}},\
  \bibinfo {pages} {3616} (\bibinfo {year} {1989})}\BibitemShut {NoStop}%
\bibitem [{\citenamefont {Bl\"{o}chl}\ \emph {et~al.}(1994)\citenamefont
  {Bl\"{o}chl}, \citenamefont {Jepsen},\ and\ \citenamefont
  {Andersen}}]{blochl_improved_1994}%
  \BibitemOpen
  \bibfield  {author} {\bibinfo {author} {\bibfnamefont {P.}~\bibnamefont
  {Bl\"{o}chl}}, \bibinfo {author} {\bibfnamefont {O.}~\bibnamefont {Jepsen}},
  \ and\ \bibinfo {author} {\bibfnamefont {O.}~\bibnamefont {Andersen}},\
  }\href {\doibase 10.1103/PhysRevB.49.16223} {\bibfield  {journal} {\bibinfo
  {journal} {Phys. Rev. B}\ }\textbf {\bibinfo {volume} {49}},\ \bibinfo
  {pages} {16223} (\bibinfo {year} {1994})}\BibitemShut {NoStop}%
\bibitem [{\citenamefont {Marlo}\ and\ \citenamefont
  {Milman}(2000)}]{marlo_density-functional_2000}%
  \BibitemOpen
  \bibfield  {author} {\bibinfo {author} {\bibfnamefont {M.}~\bibnamefont
  {Marlo}}\ and\ \bibinfo {author} {\bibfnamefont {V.}~\bibnamefont {Milman}},\
  }\href {\doibase 10.1103/PhysRevB.62.2899} {\bibfield  {journal} {\bibinfo
  {journal} {Phys. Rev. B}\ }\textbf {\bibinfo {volume} {62}},\ \bibinfo
  {pages} {2899} (\bibinfo {year} {2000})}\BibitemShut {NoStop}%
\bibitem [{\citenamefont {Wang}\ \emph {et~al.}(1998)\citenamefont {Wang},
  \citenamefont {Weiss}, \citenamefont {Shaikhutdinov}, \citenamefont {Ritter},
  \citenamefont {Petersen}, \citenamefont {Wagner}, \citenamefont
  {Schl\"{o}gl},\ and\ \citenamefont {Scheffler}}]{wang_hematite_1998}%
  \BibitemOpen
  \bibfield  {author} {\bibinfo {author} {\bibfnamefont {X.-G.}\ \bibnamefont
  {Wang}}, \bibinfo {author} {\bibfnamefont {W.}~\bibnamefont {Weiss}},
  \bibinfo {author} {\bibfnamefont {S.}~\bibnamefont {Shaikhutdinov}}, \bibinfo
  {author} {\bibfnamefont {M.}~\bibnamefont {Ritter}}, \bibinfo {author}
  {\bibfnamefont {M.}~\bibnamefont {Petersen}}, \bibinfo {author}
  {\bibfnamefont {F.}~\bibnamefont {Wagner}}, \bibinfo {author} {\bibfnamefont
  {R.}~\bibnamefont {Schl\"{o}gl}}, \ and\ \bibinfo {author} {\bibfnamefont
  {M.}~\bibnamefont {Scheffler}},\ }\href {\doibase
  10.1103/PhysRevLett.81.1038} {\bibfield  {journal} {\bibinfo  {journal}
  {Phys. Rev. Lett.}\ }\textbf {\bibinfo {volume} {81}},\ \bibinfo {pages}
  {1038} (\bibinfo {year} {1998})}\BibitemShut {NoStop}%
\bibitem [{\citenamefont {Reuter}\ and\ \citenamefont
  {Scheffler}(2001)}]{reuter_composition_2001}%
  \BibitemOpen
  \bibfield  {author} {\bibinfo {author} {\bibfnamefont {K.}~\bibnamefont
  {Reuter}}\ and\ \bibinfo {author} {\bibfnamefont {M.}~\bibnamefont
  {Scheffler}},\ }\href {\doibase 10.1103/PhysRevB.65.035406} {\bibfield
  {journal} {\bibinfo  {journal} {Phys. Rev. B}\ }\textbf {\bibinfo {volume}
  {65}},\ \bibinfo {pages} {035406} (\bibinfo {year} {2001})}\BibitemShut
  {NoStop}%
\bibitem [{\citenamefont {Lee}\ \emph {et~al.}(2011)\citenamefont {Lee},
  \citenamefont {Behera}, \citenamefont {Wachsman}, \citenamefont {Phillpot},\
  and\ \citenamefont {Sinnott}}]{lee_stoichiometry_2011}%
  \BibitemOpen
  \bibfield  {author} {\bibinfo {author} {\bibfnamefont {C.-W.}\ \bibnamefont
  {Lee}}, \bibinfo {author} {\bibfnamefont {R.}~\bibnamefont {Behera}},
  \bibinfo {author} {\bibfnamefont {E.}~\bibnamefont {Wachsman}}, \bibinfo
  {author} {\bibfnamefont {S.}~\bibnamefont {Phillpot}}, \ and\ \bibinfo
  {author} {\bibfnamefont {S.}~\bibnamefont {Sinnott}},\ }\href {\doibase
  10.1103/PhysRevB.83.115418} {\bibfield  {journal} {\bibinfo  {journal} {Phys.
  Rev. B}\ }\textbf {\bibinfo {volume} {83}},\ \bibinfo {pages} {115418}
  (\bibinfo {year} {2011})}\BibitemShut {NoStop}%
\bibitem [{\citenamefont {Wang}\ \emph {et~al.}(2010)\citenamefont {Wang},
  \citenamefont {Dai}, \citenamefont {Gao}, \citenamefont {Ruan}, \citenamefont
  {Wang},\ and\ \citenamefont {Sun}}]{wang_surface_2010}%
  \BibitemOpen
  \bibfield  {author} {\bibinfo {author} {\bibfnamefont {C.}~\bibnamefont
  {Wang}}, \bibinfo {author} {\bibfnamefont {Y.}~\bibnamefont {Dai}}, \bibinfo
  {author} {\bibfnamefont {H.}~\bibnamefont {Gao}}, \bibinfo {author}
  {\bibfnamefont {X.}~\bibnamefont {Ruan}}, \bibinfo {author} {\bibfnamefont
  {J.}~\bibnamefont {Wang}}, \ and\ \bibinfo {author} {\bibfnamefont
  {B.}~\bibnamefont {Sun}},\ }\href {\doibase 10.1016/j.ssc.2010.04.034}
  {\bibfield  {journal} {\bibinfo  {journal} {Solid State Commun.}\ }\textbf
  {\bibinfo {volume} {150}},\ \bibinfo {pages} {1370} (\bibinfo {year}
  {2010})}\BibitemShut {NoStop}%
\bibitem [{\citenamefont {Neugebauer}\ and\ \citenamefont
  {Scheffler}(1992)}]{neugebauer_adsorbate-substrate_1992}%
  \BibitemOpen
  \bibfield  {author} {\bibinfo {author} {\bibfnamefont {J.}~\bibnamefont
  {Neugebauer}}\ and\ \bibinfo {author} {\bibfnamefont {M.}~\bibnamefont
  {Scheffler}},\ }\href {\doibase 10.1103/PhysRevB.46.16067} {\bibfield
  {journal} {\bibinfo  {journal} {Phys. Rev. B}\ }\textbf {\bibinfo {volume}
  {46}},\ \bibinfo {pages} {16067} (\bibinfo {year} {1992})}\BibitemShut
  {NoStop}%
\bibitem [{\citenamefont {Wyckoff}(1971)}]{wyckoff_crystal_1971}%
  \BibitemOpen
  \bibfield  {author} {\bibinfo {author} {\bibfnamefont {R.}~\bibnamefont
  {Wyckoff}},\ }\href@noop {} {\emph {\bibinfo {title} {Crystal Structures}}},\
  \bibinfo {edition} {2nd}\ ed.\ (\bibinfo  {publisher} {Interscience, New
  York},\ \bibinfo {year} {1971})\BibitemShut {NoStop}%
\bibitem [{\citenamefont {Reguzzoni}\ \emph {et~al.}(2012)\citenamefont
  {Reguzzoni}, \citenamefont {Fasolino}, \citenamefont {Molinari},\ and\
  \citenamefont {Righi}}]{reguzzoni12}%
  \BibitemOpen
  \bibfield  {author} {\bibinfo {author} {\bibfnamefont {M.}~\bibnamefont
  {Reguzzoni}}, \bibinfo {author} {\bibfnamefont {A.}~\bibnamefont {Fasolino}},
  \bibinfo {author} {\bibfnamefont {E.}~\bibnamefont {Molinari}}, \ and\
  \bibinfo {author} {\bibfnamefont {M.~C.}\ \bibnamefont {Righi}},\ }\href
  {\doibase 10.1103/PhysRevB.86.245434} {\bibfield  {journal} {\bibinfo
  {journal} {Phys. Rev. B}\ }\textbf {\bibinfo {volume} {86}},\ \bibinfo
  {pages} {245434} (\bibinfo {year} {2012})}\BibitemShut {NoStop}%
\bibitem [{\citenamefont {Stoyanov}\ \emph {et~al.}(2014)\citenamefont
  {Stoyanov}, \citenamefont {Romero}, \citenamefont {Merz}, \citenamefont
  {Kopnarski}, \citenamefont {Stricker}, \citenamefont {Stemmer}, \citenamefont
  {Dienwiebel},\ and\ \citenamefont {Moseler}}]{stoyanov_nanoscale_2014}%
  \BibitemOpen
  \bibfield  {author} {\bibinfo {author} {\bibfnamefont {P.}~\bibnamefont
  {Stoyanov}}, \bibinfo {author} {\bibfnamefont {P.}~\bibnamefont {Romero}},
  \bibinfo {author} {\bibfnamefont {R.}~\bibnamefont {Merz}}, \bibinfo {author}
  {\bibfnamefont {M.}~\bibnamefont {Kopnarski}}, \bibinfo {author}
  {\bibfnamefont {M.}~\bibnamefont {Stricker}}, \bibinfo {author}
  {\bibfnamefont {P.}~\bibnamefont {Stemmer}}, \bibinfo {author} {\bibfnamefont
  {M.}~\bibnamefont {Dienwiebel}}, \ and\ \bibinfo {author} {\bibfnamefont
  {M.}~\bibnamefont {Moseler}},\ }\href {\doibase
  10.1016/j.actamat.2013.12.029} {\bibfield  {journal} {\bibinfo  {journal}
  {Acta Mater.}\ }\textbf {\bibinfo {volume} {67}},\ \bibinfo {pages} {395}
  (\bibinfo {year} {2014})}\BibitemShut {NoStop}%
\bibitem [{\citenamefont {Momma}\ and\ \citenamefont
  {Izumi}(2011)}]{momma_vesta_2011}%
  \BibitemOpen
  \bibfield  {author} {\bibinfo {author} {\bibfnamefont {K.}~\bibnamefont
  {Momma}}\ and\ \bibinfo {author} {\bibfnamefont {F.}~\bibnamefont {Izumi}},\
  }\href {\doibase 10.1107/S0021889811038970} {\bibfield  {journal} {\bibinfo
  {journal} {J. Appl. Crystallogr.}\ }\textbf {\bibinfo {volume} {44}},\
  \bibinfo {pages} {1272} (\bibinfo {year} {2011})}\BibitemShut {NoStop}%
\bibitem [{\citenamefont {Childs}\ \emph {et~al.}(2012)\citenamefont {Childs},
  \citenamefont {Brugger}, \citenamefont {Whitlock}, \citenamefont {Meredith},
  \citenamefont {Ahern}, \citenamefont {Pugmire}, \citenamefont {Biagas},
  \citenamefont {Miller}, \citenamefont {Harrison}, \citenamefont {Weber},
  \citenamefont {Krishnan}, \citenamefont {Fogal}, \citenamefont {Sanderson},
  \citenamefont {Garth}, \citenamefont {Bethel}, \citenamefont {Camp},
  \citenamefont {R\"{u}bel}, \citenamefont {Durant}, \citenamefont {Favre},\
  and\ \citenamefont {Navr\'{a}til}}]{childs_visit:_2012}%
  \BibitemOpen
  \bibfield  {author} {\bibinfo {author} {\bibfnamefont {H.}~\bibnamefont
  {Childs}}, \bibinfo {author} {\bibfnamefont {E.}~\bibnamefont {Brugger}},
  \bibinfo {author} {\bibfnamefont {B.}~\bibnamefont {Whitlock}}, \bibinfo
  {author} {\bibfnamefont {J.}~\bibnamefont {Meredith}}, \bibinfo {author}
  {\bibfnamefont {S.}~\bibnamefont {Ahern}}, \bibinfo {author} {\bibfnamefont
  {D.}~\bibnamefont {Pugmire}}, \bibinfo {author} {\bibfnamefont
  {K.}~\bibnamefont {Biagas}}, \bibinfo {author} {\bibfnamefont
  {M.}~\bibnamefont {Miller}}, \bibinfo {author} {\bibfnamefont
  {C.}~\bibnamefont {Harrison}}, \bibinfo {author} {\bibfnamefont {G.~H.}\
  \bibnamefont {Weber}}, \bibinfo {author} {\bibfnamefont {H.}~\bibnamefont
  {Krishnan}}, \bibinfo {author} {\bibfnamefont {T.}~\bibnamefont {Fogal}},
  \bibinfo {author} {\bibfnamefont {A.}~\bibnamefont {Sanderson}}, \bibinfo
  {author} {\bibfnamefont {C.}~\bibnamefont {Garth}}, \bibinfo {author}
  {\bibfnamefont {E.}~\bibnamefont {Bethel}}, \bibinfo {author} {\bibfnamefont
  {D.}~\bibnamefont {Camp}}, \bibinfo {author} {\bibfnamefont {O.}~\bibnamefont
  {R\"{u}bel}}, \bibinfo {author} {\bibfnamefont {M.}~\bibnamefont {Durant}},
  \bibinfo {author} {\bibfnamefont {J.}~\bibnamefont {Favre}}, \ and\ \bibinfo
  {author} {\bibfnamefont {P.}~\bibnamefont {Navr\'{a}til}},\ }in\ \href@noop
  {} {\emph {\bibinfo {booktitle} {High Performance {Visualization--Enabling}
  Extreme-Scale Scientific Insight}}}\ (\bibinfo {year} {2012})\ pp.\ \bibinfo
  {pages} {357--372}\BibitemShut {NoStop}%
\bibitem [{\citenamefont {Humphrey}\ \emph {et~al.}(1996)\citenamefont
  {Humphrey}, \citenamefont {Dalke},\ and\ \citenamefont
  {Schulten}}]{humphrey_vmd:_1996}%
  \BibitemOpen
  \bibfield  {author} {\bibinfo {author} {\bibfnamefont {W.}~\bibnamefont
  {Humphrey}}, \bibinfo {author} {\bibfnamefont {A.}~\bibnamefont {Dalke}}, \
  and\ \bibinfo {author} {\bibfnamefont {K.}~\bibnamefont {Schulten}},\ }\href
  {\doibase 10.1016/0263-7855(96)00018-5} {\bibfield  {journal} {\bibinfo
  {journal} {J. Mol. Graphics}\ }\textbf {\bibinfo {volume} {14}},\ \bibinfo
  {pages} {33} (\bibinfo {year} {1996})}\BibitemShut {NoStop}%
\end{thebibliography}%
%\end{thebibliography}

\end{document}